\documentclass[a4paper,11pt]{article}
\pdfoutput=1 % if your are submitting a pdflatex (i.e. if you have
\usepackage{jcappub} % for details on the use of the package, please
\usepackage[T1]{fontenc} % if needed

\usepackage{graphicx}
\usepackage{xcolor}
\usepackage{amsmath,amssymb}
\usepackage{bm}
\usepackage{hyperref}
\usepackage{url}
\usepackage{color}

\newcommand{\be}{\begin{equation}}
\newcommand{\ee}{\end{equation}}

\newcommand{\de}{{\rm d}}
\newcommand{\hi}{\textsc{Hi}}

\newcommand{\sigmav}{\langle \sigma v \rangle}
\newcommand{\fermi}{{\it Fermi}-LAT}
\newcommand{\HI}{{\rm HI}}

\title{\boldmath Synergies across the spectrum for particle dark matter indirect detection: how HI  intensity mapping meets gamma rays}

\author[a,b,c]{Elena Pinetti}
\author[a,b,d]{Stefano Camera}
\author[a,b]{Nicolao Fornengo}
\author[a,b]{Marco Regis}

\affiliation[a]{Dipartimento di Fisica, Universit\`a di Torino, Via P.\ Giuria 1, 10125 Torino, Italy}
\affiliation[b]{INFN, Sezione di Torino, Via P.\ Giuria 1, 10125 Torino, Italy}
\affiliation[c]{LPTHE, Sorbonne Université,
4 Place Jussieu, F-75252 Paris, France}
\affiliation[d]{INAF, Osservatorio Astrofisico di Torino, Strada Osservatorio 20, 10025 Pino Torinese, Italy}

\emailAdd{elena.pinetti@unito.it}
\emailAdd{stefano.camera@unito.it}
\emailAdd{nicolao.fornengo@unito.it}
\emailAdd{marco.regis@unito.it}

\abstract{Neutral hydrogen (\hi) intensity mapping traces the large-scale distribution of matter in the Universe and therefore should correlate with the gamma-ray emission originated from particle dark matter annihilation or from active galactic nuclei and star-forming galaxies, since the related processes occur in the same cosmic structures hosting \hi. In this paper, we derive the cross-correlation signal between the brightness temperature of the 21-cm line emission of the \hi\ spin-flip transition in the Universe and the unresolved gamma-ray background. Specifically, we derive forecasts for the cross-correlation signal by focussing on the opportunities offered by the combination of the {\it Fermi}-Large Area Telescope (LAT) gamma-ray sensitivity with the expectations of the \hi\ intensity mapping measurements from future radio telescopes, for which we concentrate on the Square Kilometre Array (SKA) and MeerKAT, one of its precursors. We find that the combination of MeerKAT with the current  \fermi\ statistics has the potential to provide a first hint of the cross-correlation signal originated by astrophysical sources, with a signal-to-noise ratio (SNR) of 3.7. With SKA Phase 1 and SKA Phase 2, the SNR is predicted to increase up to 5.7 and 8.2, respectively. The bounds on dark matter properties attainable with SKA combined with the current statistics of \fermi\ are predicted to be comparable to those obtained from other techniques able to explore the unresolved components of the gamma-ray background. The enhanced capabilities of SKA Phase 2, combined with a future generation gamma-ray telescope with improved specifications, can allow us to investigate the whole mass window for weakly interacting massive particles up to the TeV scale.
}

\begin{document}
\maketitle
\flushbottom

\section{Introduction}
\label{sec:intro}

Cross-correlations between gravitational tracers of the large-scale distribution of matter in the Universe and the electromagnetic cosmic backgrounds have been proposed as a promising tool to explore the origin of the unresolved components of these radiation fields---a place where an elusive particle dark matter signal might hide \cite{Camera:2012cj}. Dark matter (DM) is in fact expected to produce annihilation or decay signals, a most notable example being the gamma-ray emission produced by essentially any kind of weakly interacting massive particles (WIMP). However, those signals are remarkably faint and immersed in an overwhelming astrophysical background. By bringing three-dimensional spatial information from the gravitational tracers, the cross-correlation technique adds relevant details that can potentially assist in disentangling a DM signal from the other astrophysical emissions \cite{Camera:2014rja,Fornengo2014,Ando:2013xwa}.

The cross-correlation technique has been employed to investigate the faint end of the unresolved gamma-ray background (UGRB) by using galaxies \cite{Xia2011,Ando:2013xwa,Ando:2014aoa,Regis2015,Cuoco2015,Xia2015,Shirasaki:2015nqp,Cuoco:2017bpv,Ammazzalorso:2018evf,Hashimoto:2019obg}, clusters of galaxies \cite{Branchini:2016glc,Hashimoto:2018ztv,Colavincenzo:2019jtj,Tan:2019gmb}, lensing of the cosmic microwave background \cite{Fornengo:2014cya,Feng2016} and the weak lensing effect of cosmic shear \cite{Camera:2012cj,Camera:2014rja,Shirasaki2014,Troster:2016sgf,Shirasaki:2016kol,Shirasaki:2018dkz,Ammazzalorso:2019wyr}. In this paper we investigate a novel possibility, that will become available in the next years with the deep investigation of the 21-cm emission from cosmic neutral hydrogen (\hi), explored through the intensity mapping technique and made possible with the oncoming generation of radio telescopes, most notably the Square Kilometre Array (SKA). The 21-cm emission works as a probe of the underlying matter field and, due to the large-scale structure of the Universe, it is intrinsically anisotropic. If DM is composed by a new kind of elementary particle, able to produce a faint  radiation by means of its self-annihilation, then this radiation traces the same DM structure probed by \hi\ and shares a statistically common pattern of fluctuations. (Note that DM decay is also a viable scenario, provided the decay is sufficiently suppressed in order to allow the particle to be long-lived on cosmological timescales.) Emission from unresolved astrophysical sources, like active galactic nuclei or star-forming galaxies, also contribute to the cosmic radiation with a pattern that is necessarily correlated with the gravitational tracer (here represented by the 21-cm line), as both are hosted by the same DM haloes.

For this reason, a level of cross-correlation between the \hi\ brightness temperature with those cosmic radiation fields is expected. We study this cross-correlation signal by focussing on the high-energy tail of the cosmic radiation, namely on the broad gamma-ray band, which is relevant both for the astrophysical sources and for particle DM in terms of WIMPs, which have the ability to produce gamma rays through their annihilation products, like final state radiation of produced leptons or decays of hadrons generated in the annihilation process.

The tool to investigate the cross-correlation signal is the angular power spectrum of the correlation between the fluctuations of the two fields, namely 21-cm and UGRB. We quantify the size of the signal and derive forecasts on the ability to detect this signal by adopting a full-sky, large field-of view gamma-ray telescope like the {\it Fermi} Large-Area Telescope (LAT), combined with the intensity mapping observations of SKA \citep{Santos:2015bsa,Bacon:2018dui} and, on a shorter timescale, its precursor MeerKAT \citep{Santos:2017qgq}. We show that indeed the cross-correlation signal is potentially detectable with the SKA, with a possible hint attainable already with MeerKAT, and we derive the prospects to set bounds on (or detect a signal for) the relevant DM particle physics properties---namely its mass and annihilation cross-section. Finally, we quantify what prospects can be achieved in the long term by the Phase 2 of SKA combined with a future generation gamma-ray telescope with larger exposure and improved angular resolution.

\section{The cross-correlation signal}
\label{sec:signal}
The angular power spectrum (APS) of the cross-correlation signal between two observables $i$ and $j$ can be written as \cite{cooray,Fornengo2014}:
\be
 C_l^{ij} = \int \frac{\de\chi}{\chi^2} \, W_i (\chi) W_j(\chi) P_{ij}\left(k=\frac{l}{\chi}, \chi\right),
\label{eq:APS}
\ee
where $l$ is an angular multipole. % \stef{where $\chi(z)$ is the radial comoving distance to redshift $z$, such that $\de\chi/\de z=1/H(z)$ with $H$ the Hubble rate, $W_i$ is the kernel of the $i$th observable, and $P_{ij}$ is the cross-power spectrum of the two cosmological fields seeding the two observables.}
In our main case of study in this paper, $i$ and $j$ will respectively label the \hi\ brightness temperature $T_{\rm b}$ and the gamma-ray intensity $I_\gamma$. 
%$\langle I_i \rangle$ is the average value of the source field $i$
The quantity denoted by $\chi(z)$ is the radial comoving distance to redshift $z$, for which we have $\de\chi = c\de z/H(z)$ (in a flat Universe) with $H(z)=H_0 [\Omega_{\rm m} (1+z)^3+\Omega_{\Lambda}]^{1/2}$ the Hubble rate at redshift $z$, $\Omega_{\rm m}$ and $\Omega_{\Lambda}$ the matter and dark energy density parameters, and $H_0\equiv H(z=0)$ the Hubble constant. The window function $W_{i}(\chi)$ contains the information on how the observable $i$ is distributed in redshift and its shape strongly depends on the signal under consideration. Lastly, $P_{ij}$ is the three-dimensional Fourier-space cross-power spectrum of the fluctuations $f_i = g_i - \langle g \rangle$ of the density fields $g_i$ of the two obervables $i$ and $j$ ($\langle g_i \rangle$ is the average of the field $g_i$). It is implicitly defined as:
\begin{equation}
\langle \tilde f_i(\chi,{\bm k}) \tilde f_j(\chi,{\bm k}') \rangle = (2\pi)^2 \delta^3({\bm k} - {\bm k}')  P_{ij}(k,\chi),
\end{equation}
where tilde denotes Fourier transform and the density fields are related to the source fields via:
\begin{equation}
I_i(\vec n) = \int \de\chi\,g_i(\chi, \vec n) \tilde W_i(\chi)
\end{equation}
with $\vec n$ referring to the line-of-sight direction in the sky. The window functions are normalised through $ W_i(\chi)=\langle g_i(\chi)\rangle \tilde W_i(\chi)$ such that $\langle I_i \rangle = \int \de\chi W_i(\chi)$. In Eq.~(\ref{eq:APS}) we adopt the Limber approximation \cite{Limber1953,Kaiser1992}, valid for $l\gg1$, which sets $k=l/\chi$. We closely follow Refs.~\cite{Fornengo2014,Camera:2014rja} for notations and definitions.

In the halo model \cite{cooray}, the power spectrum at a given redshift can be separated into two terms, referred to as the 1-halo and the 2-halo term, i.e.\ :
\begin{equation}
P(k) = P^{\rm 1h} (k) + P^{\rm 2h}(k).
\label{eq:pk}
\end{equation}
The general expressions for $P^{\rm 1h}$ and $P^{\rm 2h}$ are:
\begin{align}
P_{ij}^{\rm 1h} (k) &= \int_{M_{\rm min}}^{M_{\rm max}} \de M \frac{\de N}{\de M} \tilde f_i^*(k|M) \tilde f_j(k|M), \label{eqn:P1h} \\
P_{ij}^{\rm 2h} (k) &= \left [\int_{M_{\rm min}}^{M_{\rm max}} \de M_1  \frac{\de N}{\de M_1} \,b(M_1) \tilde f_i^*(k|M_1) \right] \left[\int_{M_{\rm min}}^{M_{\rm max}} \de M_2  \frac{\de N}{\de M_2}\, b(M_2) \tilde f_j(k|M_2) \right] P_{\rm lin} (k),
\label{eqn:P2h}
\end{align}
where $M$ denotes the mass of the halo, $\de N/\de M$ is the halo mass function, $b(M)$ is the bias of the DM haloes with respect to matter, and $P_{\rm lin}$ is the linear matter power spectrum at redshift $z$. Eqs~(\ref{eq:pk}), (\ref{eqn:P1h}) and (\ref{eqn:P2h}) contain a redshift dependence arising from all its components, but it has been omitted to ease the notation. $M_{\rm min}$ and $M_{\rm max}$ denote the minimum and maximum halo mass,  for which we assume $M_{\rm min}=10^{-6} \,M_{\odot}$, corresponding to a typical reference value for the minimal halo mass for DM composed by WIMPs, and $M_{\rm min}=10^{18} \,M_{\odot}$, above super-cluster size. (Note that the choice of $M_{\rm max}$ does not sensibly affect the results.) For the halo mass function and the linear halo bias, we adopt the prescriptions of Ref.~\cite{sheth_tormen} and Ref.~\cite{Sheth:1999mn}, respectively. 

The specific expression for the power spectrum of the cross-correlation between the 21-cm brightness temperature and the gamma-ray intensity field produced by DM or by astrophysical sources will be specified explicitly below. We just notice here that in case of astrophysical sources, they are better characterised by their luminosity rather than mass of the hosting halo: in this case we need to replace the halo mass $M$ by the luminosity $L$ and the halo mass function $\de N/\de M(M,z)$ by their gamma-ray luminosity function (GLF) $\de N/\de L \equiv \phi(L,z)$. For the range of angular scales probed by the cross-correlation APS, we will consider astrophysical sources as point sources. The \hi\ brightness temperature and the gamma-ray emission from DM instead are extended distributions: a noticeable difference to take into account is that while \hi\ traces the total matter density, gamma rays from DM annihilation trace the DM density squared, since each annihilation process involves two DM particles.

Under the hypothesis of gaussianity, the variance on the predicted cross-correlation APS is:
\begin{equation}
    (\Delta C_l^{ij})^2 = \frac{1}{(2l+1) f_{\rm sky}}  \left\{\left(C_l^{ij}\right)^2 + \left[C_l^{ii} +  \frac{N^i}{(B^i_l)^2} \right] \left[C_l^{jj} + \frac{N^j}{(B^j_l)^2} \right] \right\}
    \label{eq:deltaClcross}
\end{equation}
where $C_l^{ii}$ and $C_l^{jj}$ denote the auto-correlation APS of the two  observables, $N^i$ and $N^j$ are their corresponding noises, $B^i_l$ and $B^j_l$ their beam functions which describe the angular resolution of the detectors used for the two observables and $f_{\rm sky}$ is the observed fraction of the sky.

\section{Neutral hydrogen intensity mapping}
\label{sec:hydrogen}
\begin{table}[t]
\footnotesize
\centering
\begin{tabular}{lcccccccccc}
 \hline
 & $S\,[\deg^2]$ & $t\,[10^3\,\mathrm{hr}]$ & $N_{\rm d}$ & $D_{\rm dish}\,[\rm m]$ & $D_{\rm interf}\,[\rm km]$ & $N_{\rm b}$ & $f_{\rm sky}$ & $[z_{\rm min}, z_{\rm max}]$ & \\
\hline
MeerKAT & 4,000 & $4$ & 64 & 13.5 & $1$ & 1 & 0.097 & $[0.4, 1.45]$  & UHF-band\\
& & & & & &  & & $[0.0, 0.58]$ & L-band\\
\hline
SKA1 & 25,000 & $10$ & 133+64 & 14.5 & $3$ & 1 & 0.61 & $[0.35, 3]$ & Band 1 \\
& & & &  & & & & $[0.0, 0.5]$ & Band 2 \\
\hline
SKA2 & 30,000 & $10$ & 2,000  & 14.5 & 10 & 36 & 0.72 & $[0.35, 3]$ & Band 1 \\
& & & &  & & & & $[0.0, 0.5]$ & Band 2 \\
\hline
\end{tabular}
\caption{Technical specifications for MeerKAT, SKA1 and SKA2, as used in our analysis. $S$ denotes the area covered by the survey,  $t$ is the assumed total observation time, $N_{\rm d}$ is the number of dishes, $D_{\rm dish}$ is the dish diameter, $D_{\rm interf}$ is the interferometer mean baseline, $N_{\rm b}$ is the PAF, $f_{\rm sky}$ is the fraction of sky covered by the detector. The last columns show the bands in redshift adopted in the analysis and their assigned names.}
\label{tab:specifiche_IM}
\end{table}

In this paper, we scrutinise the opportunity to employ the 21-cm line of the hyperfine transition of \hi\ as a tracer of the DM distribution in the late, post-reionisation Universe by means of a technique known as `intensity mapping' \citep{Chang:2010jp, loeb2008, visbal2010}. It is based on the assumption that, after the end of reionisation, \hi\ only survives within (mostly blue) galaxies, in turn residing in DM haloes. On the one hand, this is not dissimilar to studies of galaxy clustering \citep{2011ApJ...736...59Z,Alam:2016hwk,Pezzotta:2016gbo,Norberg:2008tg}, where the three-dimensional distribution of galaxies is used as a proxy of the underlying cosmic large-scale structure. On the other hand, galaxy surveys are threshold experiments --- only galaxies above a certain flux are detected, whose type, number, and bias are thus implicitly influenced by the threshold itself. Moreover, only high signal-to-noise detected objects enter a galaxy catalogue, to ensure pure samples.

Oppositely, mapping the intensity of the brightness temperature of \hi\ on the sky needs not to resolve individual galaxies, but it effectively collects \textit{all} photons of a given, observed frequency. The result is a temperature map resembling in a way maps of the cosmic microwave background; in this case, though, hot pixels, where significant 21-cm radiation is seen, are located in the direction where a large number of \hi\ galaxies reside; cold pixels, instead, are related to voids or troughs in the large-scale \hi\ galaxy web.

One of the main advantages offered by the intensity mapping technique is that detection of the 21-cm line provides redshift information for free, since the observed photon frequency, $\nu_{\rm o}$, is related to the frequency at emission, $\nu_{\rm e} = 1420\,\mathrm{MHz}$, via:
\begin{equation}
\nu_{\rm o} = \frac{\nu_{\rm e}}{1+z}.
\end{equation}
As a drawback, measuring the integrated redshifted \hi\ emission of a large number of galaxies within one angular pixel does imply a poorer angular resolution, compared to galaxy surveys. This effect is encapsulated in the beam, as in Eq.~\eqref{eq:deltaClcross}.

An additional benefit is that the measured signal will be larger as we expect that there are numerous \hi\ galaxies in each three-dimensional voxel. In this regard, typical frequency and angular resolutions are, respectively, about 5 MHz and 1 degree, corresponding to a comoving volume of around $10^5\,\mathrm{Mpc}^3$. Each cell hosts $\sim10^6$ DM haloes with mass in the range $10^8 - 10^{15}\,M_{\odot}$ and around $31,000$ within the range $5\times10^9$ to $10^{12}\,M_\odot$, containing most of the \hi\ in the Universe \cite{Santos:2015bsa}.

The \hi\ density parameter is defined as:
\begin{equation}
\Omega_{\HI} (z) \equiv (1+z)^{-3} \,\frac{\overline{\rho}_{\HI}(z)}{\rho_{{\rm c},0}},
\end{equation}
where $\rho_{{\rm c},0}$ is the critical density of the Universe at the present time and  $\overline{\rho}_{\HI}(z)$ is the \hi\ mean density at redshift $z$, which depends on the total mass $M_{\HI} (M,z)$ of \hi\ within a halo of mass $M$ through:
\begin{equation}
\overline{\rho}_{\HI} (z) = \int_{M_{\rm min}}^{M_{\rm max}} \de M \,\frac{\de N}{\de M}(M,z) \,M_{\HI} (M,z).
\end{equation}
The average brightness temperature can be parameterised as \cite{Santos:2015bsa}:
%
%\begin{equation}
%\overline{T}_{\rm b} (z) \approx 566 \hspace{0.3mm} h \hspace{0.5mm} \left(\frac{H_0}{H(z)} \right) \left( \frac{\Omega_{\rm HI}(z)}{0.003} \right) \hspace{0.5mm} (1+z)^2 \hspace{1mm} \mu K
%    \label{eq:tbright}
%\end{equation}
%
%where $H_0$ is the Hubble rate at present time and $h$ is its value in units of $100$ km/s Mpc$^{-1}$.
 %
 \begin{equation}
    \overline{T}_{\rm b}(z) = 44 \, \mu K  \,\left(\frac{\Omega_{\HI}(z) \, h}{2.45 \times 10^{-4}} \right) \frac{(1+z)^2}{E(z)},
    \label{eq:tbright}
 \end{equation}
 where $E(z)= H(z)/H_0$ 
 %with $H(z)=H_0 [\Omega_{\rm m} (1+z)^3+\Omega_{\Lambda}]^\frac{1}{2}$ denoting the Hubble parameter in a flat Universe at redshift $z$, $H_0$ the Hubble constant 
 and $h$ is the Hubble constant in units of $100$ km/s Mpc$^{-1}$.

If the average signal scales linearly with DM perturbations, the brightness temperature at frequency $\nu$ and angular position $\Omega$ can be written as \cite{Santos:2015bsa}:
\begin{equation}
T_{\rm b} (\nu, \Omega) \approx \overline{T}_{\rm b} (z) \left[ 1 + b_{\HI}(z) \hspace{0.5mm} \delta_{\rm m} (z) - \frac{1}{H(z)} \hspace{0.5mm} \frac{\de\nu}{ \de s} \right],
\end{equation}
where the scale-independent \hi\ bias, $b_{\HI}(z)$, is linked to the linear bias $b(z,M)$ through the relation (valid on large scales):
\begin{eqnarray}
b_{\HI}(z) = \frac{1}{\overline{\rho}_{\HI}(z)} \int_{M_{\rm min}}^{M_{\rm max}} \de M \,\frac{\de N}{\de M}(M,z) \,M_{\HI}(M,z) \,b(z,M).
\end{eqnarray}

A relevant ingredient to determine the \hi\ density and bias is the quantity $M_{\HI}(M,z)$, i.e.\ the mass of \hi\ within a DM halo of mass $M$ at redshift $z$. We follow here the model of Refs.~\cite{dla,p1}, based on up-to-date observations and Monte Carlo simulations, which reads:
\begin{equation} 
M_{\HI} (M,z) =  \alpha\, f_{\rm H}  \,\left( \frac{M}{10^{11} \, h^{-1} \, M_\odot} \right)^\beta \exp \left[- \left(\frac{v_{{\rm c},  0}}{v_{\rm c} (M,z)} \right)^3 \right],
\end{equation}
where $M$ is in units of $h^{-1} M_{\odot}$, $\alpha=0.176$, $f_{\rm H} = (1-Y_{\rm p}) \Omega_{\rm b}/\Omega_{\rm m}$ where $Y_{\rm p}=0.24$ denotes the primordial cosmic helium mass fraction, $\Omega_{\rm b}$ and $\Omega_{\rm m}$ the baryon and matter densities, $\beta=-0.69$, and:
\begin{equation}
v_{\rm c}(M,z) \equiv \sqrt{\frac{G M}{R_{\rm vir}(M,z)}}
\end{equation}
is the \hi\ circular velocity at the virial radius in a halo of mass $M$ at redshift $z$, with $v_{{\rm c},  0}= 10^{1.61}\,\mathrm{km\,s^{-1}}$ the minimum virial velocity of a DM halo to be able to contain \hi\ \cite{dla}.

Hydrodynamical simulations \cite{sim1} suggest that hot gas follows an altered NFW profile \cite{rhoHI} with a thermal core at about 3/4 of the scale radius $r_{\rm s} = R_{\rm vir}(M,z)/c_{\HI}(M,z)$, where $c_{\HI}$ is the \hi\ concentration parameter. Such a density profile takes the form:
\begin{equation}
\rho_{\HI} \hspace{0.5mm} (r) = \frac{\rho_0 \hspace{0.5mm} r_{\rm s}^3}{(r+0.75 \, r_{\rm s}) \hspace{0.5mm} (r+r_{\rm s})^2}.
\label{eqn:rhoHI}
\end{equation}
In the equation above, $r$ is the distance from the centre of the DM halo and the normalisation $\rho_0$ is determined by requiring that the mass within the virial radius equals $M_{\rm HI}$, viz.\ :
\begin{equation}
\int_0^{R_{\rm vir}} \de r \,4 \pi \, r^2 \,\rho_{\rm HI}(r) = M_{\HI}(M,z).
\label{eqn:rho0}
\end{equation}
The \hi\ concentration parameter $c_{\HI}(M,z)$ is assumed to be similar to the DM concentration parameter, and its dependence on the halo mass is given by the fitting function of Ref.~\cite{cHI}:
\begin{eqnarray*}
c_{\HI}(M,z) = c_{\HI,0} \left( \frac{M \,h^{-1}}{10^{11} \, M_\odot} \right)^{-0.109} \frac{4}{(1+z)^\eta},
\end{eqnarray*}
where $c_{\HI,  0}=139$ and $\eta = 0.13$.

These are the ingredients we will use in Sec.~\ref{sec:results} to calculate the cross-correlation APS between 21-cm brightness temperature fluctuations and the UGRB. In order to validate the model that we adopt for the \hi\ and its power spectrum, we calculate the real-space correlation function $\xi_{\HI}(r)$ of \hi\ galaxies and compare the result with the measurement obtained with the Arecibo Legacy Fast ALFA (ALFALFA) survey catalogue, which contains about $10,150$ \hi-selected galaxies \cite{xi}. \hi\ galaxy surveys  estimate the number-weighted correlation function that, in the case of the ALFALFA data, is very mildly dependent on the \hi\ mass \cite{xi} and therefore can be approximated by the mass-weighted correlation function. The latter is the Fourier transform of the \hi\ power spectrum (in the comparison with ALFALFA data, we consider the correlation function at $z=0$) and reads:
\begin{equation}
\xi_{\HI}(r,z)= \frac{1}{2 \pi^2} \int \frac{\de k}{k} k^3 P_{\HI}(k,z) \frac{\sin kr}{kr},
\label{eqn:xi_HIHI}
\end{equation}
where the power spectrum in our model is decomposed into the 1-halo and 2-halo terms as usual, $P_{\HI} = P_{\HI}^{\rm 1h} + P_{\HI}^{\rm 2h},$ and the two terms are:
\begin{eqnarray}
P_{\HI}^{\rm 1h}(k,z) &=& \frac{1}{\overline{\rho}^2_{\HI}(z)} \int_{M_{\rm min}}^{M_{\rm max}} dM \frac{\de N}{\de M}(M,z)\, M_{\HI}^2(M,z) \, \widetilde{u}_{\HI}^2(k|M,z),  \\
P_{\HI}^{\rm 2h}(k,z) &=& \left[\frac{1}{\overline{\rho}_{\HI}(z)} \int_{M_{\rm min}}^{M_{\rm max}} \de M \frac{\de N}{\de M}(M,z) b(z,M)  M_{\HI}(M,z) \widetilde{u}_{\HI}(k|M,z) \right]^2 P_{\rm lin}(k,z),
\end{eqnarray}
where $\widetilde{u}_{\HI}(k,z)$ denotes the Fourier transform of the \hi\ density $\rho_{\HI}(r)$ truncated at the virial radius $R_{\rm vir}$, i.e.\ :
\begin{equation}
\widetilde{u}_{\HI}(k|M,z)= \frac{4 \pi}{M_{\HI}(M,z)} \int_0^{R_{\rm vir}}  \de r \hspace{0.5mm} r^2 \rho_{\HI}(r) \, \frac{\sin kr}{kr}.
\end{equation}
Note that $\widetilde{u}_{\HI}(k,z)$ is normalised to the mass of \hi\ in the halo, which implies that proper factors of $M_{\HI}^2$ need to be taken into account in the power spectrum.

\begin{figure}[t]\centering
\includegraphics[width=0.75\textwidth]{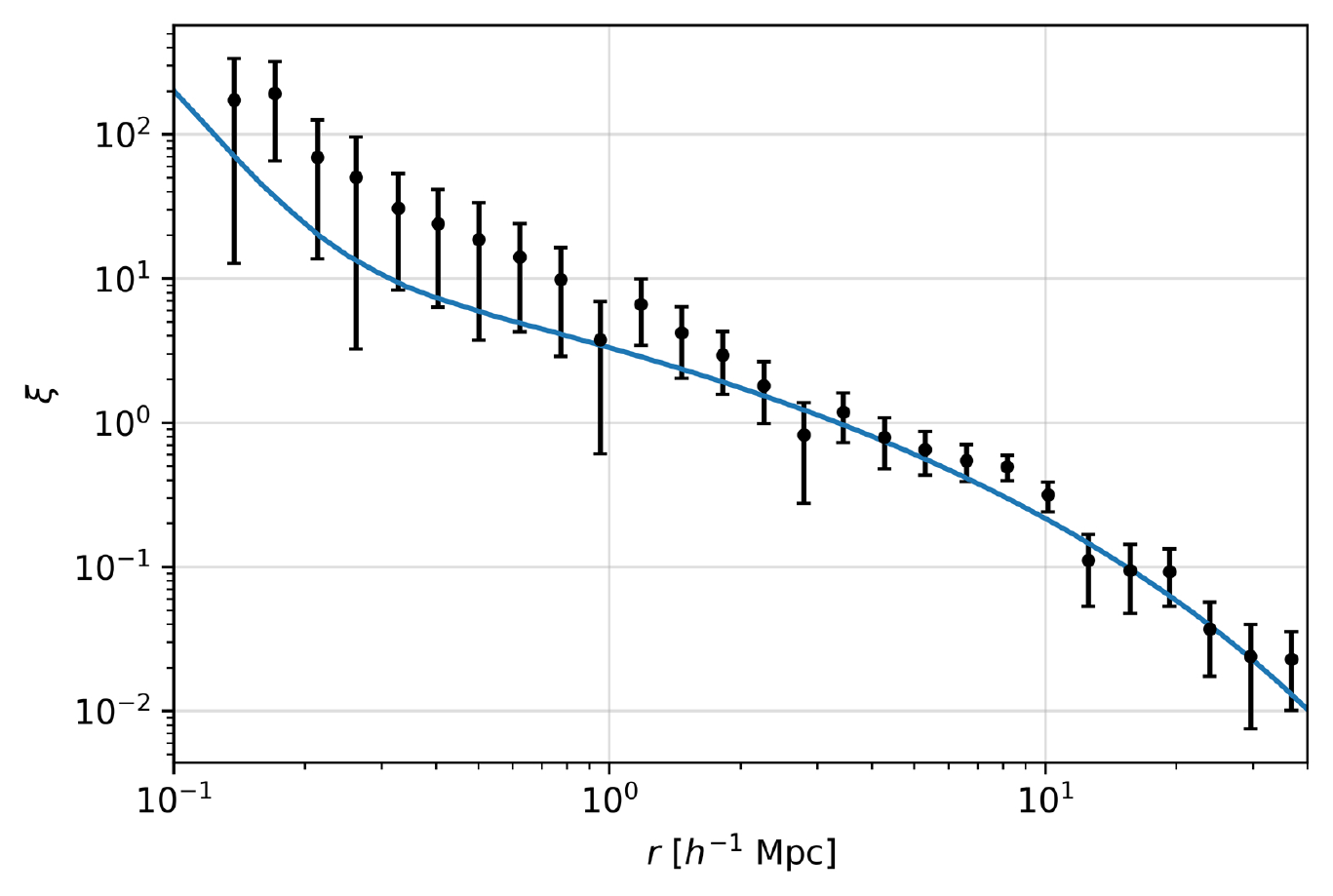}
\caption{Neutral hydrogen auto-correlation function $\xi_{\rm HI}(r)$ in physical space at $z=0$ . The line refers to the theoretical calculation obtained by using the altered NFW profile. The data points refer to the measurements of HI-selected galaxies obtained by the ALFALFA survey \cite{xi}.} 
\label{fig:xi_vs_martin}
\end{figure}

Fig.~\ref{fig:xi_vs_martin} shows the comparison between the $\xi_{\HI}$ obtained with our model and the correlation function obtained from the data of the ALFALFA survey \cite{xi}. We notice that the agreement is quite good, especially on large scales that are those of main relevance for our analysis of the cross-correlation. On small scales the model, while being in fair agreement with the data, nevertheless appears to underestimate slightly the experimental results for scales below $1\,h^{-1}\,Mpc$: the origin of this discrepancy is likely due to the imperfect knowledge of the inner parts of the \hi\ density profile on small scales, which we have modelled as in Eq.~\eqref{eqn:rhoHI} based on the results of Ref.~\cite{rhoHI}. (See also Ref.~\cite{Camera:2019iwy} for an analysis of the impact of incorrect assumptions of \hi\ clustering on cosmological parameter estimation.) We leave for a future analysis the possibility to use the \hi-selected galaxies to infer information on the \hi\ density profile.

Intensity mapping is one of the main science drivers for cosmology with the next generation of radio telescopes currently under construction, notably the SKA \citep{Maartens:2015mra,Santos:2015bsa}. As of now, some of the SKA precursors aim at dedicating observing time to intensity mapping studies, above all the Meer-Karoo Array Telescope (MeerKAT; see Ref.~\citep{Santos:2017qgq}), but it is worth naming also purpose-built instruments such as: the Canadian Hydrogen Intensity Mapping Experiment\footnote{https://chime-experiment.ca/} (CHIME); BAO In Neutral Gas Observations (BINGO; \citep{Battye:2012fd}); TianLai \citep{Chen:2012xu}; or the Five hundred metre Aperture Spherical Telescope (FAST; \citep{Smoot:2014oia}). Intensity mapping surveys can be divided into two kinds, depending on whether they will operate the system in single-dish mode or in interferometry \citep{camera}. The first set-up employs the auto-correlation signal from one or more dishes, and is best suited to study correlations at large angular separations. Oppositely, the interferometer set-up employs the cross-correlation signal from different array elements and it estimates the Fourier transform modes of the sky with high angular resolution. The most promising new generation radio telescope is certainly the SKA, which will be the world's largest radio telescope.
%, capable of detecting astronomical objects up to redshift 27. 
Phase 1 of the SKA (SKA1) will cover a frequency range from 50 MHz to 14 GHz and will be arranged in two independent sub-arrays known as SKA1-LOW (the low-frequency instrument, in Australia), and SKA1-MID (operating at mid-frequencies, in the Karoo desert, South Africa). Specifically, SKA1-MID will observe frequencies higher than 350 MHz, corresponding to the late Universe at redshift below 3; SKA1-MID can be used both in single-dish and in interferometer mode.

For our analysis of cross-correlations, we adopt a window function that follows the 21-cm brightness temperature $T_{\rm b}(z)$ of Eq.~(\ref{eq:tbright}), selected in a specific redshift range corresponding to specific radio bands \cite{battye}, namely:
 \begin{equation}
     W_{\HI}(z) = W_0(z)\,T_{\rm b} (z),
     \label{eqn:WHI}
 \end{equation}
 with:
 \begin{equation} 
      W_0(z) = \frac{\theta(z-z_{\rm min})\theta(z_{\rm max}- z)}{z_{\rm max} - z_{\rm min}},
\end{equation}
and where $z_{\rm min}$ and $z_{\rm max}$ respectively represent the lower and upper edges of the redshift bin under consideration, and $\theta$ is the Heaviside step function.

The detector beam window function of the radio telescope can be parameterised as \cite{battye}
\begin{equation}
    B^\HI_l= \exp \left[-\frac{l^2}{2} \left(\frac{1.22}{\sqrt{8  \ln2}}{\frac{\lambda_{\rm o}}{D}}\right)^2 \,\right],
\label{eq:beam}
\end{equation}
where $\lambda_{\rm o}$ is the observed wavelength of the \hi\ line, related to the wavelength of emission $\lambda_{\rm e}$ via $\lambda_{\rm o}= \lambda_{\rm e}(1+z) = 0.21$ m $(1+z)$, whereas $D$ is a reference length, which for single-dish surveys is the diameter of a dish $D_{\rm dish}$, while for interferometers we will consider the length of the core baseline $D_{\rm interf}$, which in the case of SKA1 contains approximately 75\% of the total number of dishes. 

For what concerns the noise, for a single-dish survey we follow Ref.~\cite{camera} and adopt:
\begin{equation}
    N_{\rm dish}^{\HI}= \frac{T^2_{\rm sys} \, S}{N_{\rm d} \,t\, \Delta \nu \, N_{\rm b} \, N_{\rm pol}\, \eta^2},
\end{equation}
where $T_{\rm sys}$ is the total system temperature, for which we take $T_{\rm sys}= 30 + 60 \,(300 \, \text{MHz}/\nu)^{2.55}$ K in all configurations, $S$ is the survey area, $N_{\rm d}$ is the number of dishes, $t$ is the observation time, $\Delta\nu$ is the frequency band corresponding to the redshift bin considered, $N_{\rm pol}$ is the number of polarisation states (which, for definiteness, we take equal to 2), and $\eta$ is the efficiency (assumed to be unity). Finally, $N_{\rm b}$ denotes the number of simultaneous beams that will be different from 1 when considering the use of Phased Array Feeds (PAFs) for the SKA2. 

For an interferometer survey, the noise can be written as \cite{Bull:2014rha}:
 \begin{equation}
    N_{\rm interf}^{\HI}= \frac{T^2_{\rm sys} \, S\,\,{\rm FoV}}{n(u) \, t\, \Delta \nu \, N_{\rm b}\, N_{\rm pol}\, \eta^2},
    \label{eq:noiseint}
\end{equation}
where the average number density of baselines is taken to be $n(u)=0.005$ for SKA and a factor of 10 smaller for MeerKAT \cite{Bull:talk}.  
The field of view is ${\rm FoV}\simeq \lambda_{\rm o}^2/D_{\rm dish}^2$.
Eq.~(\ref{eq:noiseint}) is valid only for angular scales smaller than $\lambda_{\rm o}/D_{\rm short}$, where the shortest baseline $D_{\rm short}$ of the array is typically a few times $D_{\rm dish}$. For definiteness, we will adopt $D_{\rm short} = 2D_{\rm dish}$, which reflects in a minimal multipole $l_{\rm cut} = \pi D_{\rm short}/(1.22 \lambda_{\rm o})$  and we will consider only $l \geq l_{\rm cut}$ for the interferometric case.  At very small angular scales (large multipoles), we account for the detector beam window function described in Eq.~(\ref{eq:beam}) which makes the noise growing exponentially.

Table ~\ref{tab:specifiche_IM} lists the instrumental specifications adopted in our analyses for MeerKAT, SKA1 and SKA2.

\section{Gamma rays}
\label{sec:gamma}
\begin{table}[t]
\centering
\begin{tabular}{ccccccc}
 \hline
Bin & $E_{\rm min}\,[\mathrm{GeV}]$ & $E_{\rm max}\,[\mathrm{GeV}]$ & $N^\gamma\,[\mathrm{cm^{-4}\,s^{-2}\,sr^{-1}}]$ & $f_{\rm sky}$ & $\sigma^{\rm Fermi}_0\,[\mathrm{deg}]$ & $E_{\rm b}\,[\mathrm{GeV}]$\\
% ~& (GeV) & (GeV) & (cm$^{-4}$ s$^{-2}$ sr$^{-1}$) & ~ & (deg) & (GeV)\\
\hline
1 & 0.5 & 1.0 & $1.056 \times 10^{-17}$ & 0.134 &  0.87 &  0.71 \\
2 & 1.0 & 1.7 & $3.548 \times 10^{-18}$ & 0.184 &  0.50  &  1.30 \\
3 & 1.7 & 2.8 & $1.375 \times 10^{-18}$ & 0.398 & 0.33  &  2.18\\
4 & 2.8 & 4.8 & $8.324 \times 10^{-19}$ & 0.482 & 0.22  &  3.67\\
5 & 4.8 & 8.3 & $3.904 \times 10^{-19}$ & 0.549 &  0.15  &  6.31\\
6 & 8.3 & 14.5 & $1.768 \times 10^{-19}$ & 0.574 & 0.11  &  11.0\\
7 & 14.5 & 22.9 & $6.899 \times 10^{-20}$ & 0.574 &  0.09  &  18.2\\
8 & 22.9 & 39.8 & $3.895 \times 10^{-20}$ &0.574 &  0.07  &  30.2\\
9 & 39.8 & 69.2 & $1.576 \times 10^{-20}$ & 0.574 &0.07  &  52.5\\
10 & 69.2 & 120.2 & $6.205 \times 10^{-21}$ & 0.574 & 0.06  &  91.2\\
11 & 120.2 & 331.1 & $3.287 \times 10^{-21}$ & 0.597 & 0.06 &  199.5\\
12 & 331.1 & 1000. & $5.094 \times 10^{-22}$ & 0.597 & 0.06 &  575.4\\
\hline
\end{tabular}
\caption{Gamma-ray energy bins adopted in the analysis, the corresponding auto-correlation noise $N^\gamma$, the fraction of sky outside the combined Galactic and point-source masks $f_{\rm sky}$ and the 68\% containment angle $\sigma^{\rm Fermi}_0$ of the \fermi\ PSF. The numbers adhere to those obtained for the UGRB auto-correlation analysis of Ref.~\cite{Ackermann:2018wlo}, which refers to 8 years of Fermi-LAT Pass 8 data products. The containment angles refer to the geometric centre of the energy bin $E_{\rm b} = (E_{\rm min} E_{\rm max})^{1/2}$ (shown in the last column)  and to the PSF2 event types, for definiteness.}
\label{tab:noise_gamma}
\end{table}

Gamma-ray astrophysics has seen a tremendous development in the last 15 years, with several thousands of sources detected, a fact that allowed us to make relevant progress in the understanding of the most violent phenomena in the Universe. However, what remains to be understood, i.e.\ the UGRB, not only incorporates relevant information on the faint and still-unexplored side of the high energy sources, but can also conceal a long-sought-after signal originated by DM composed by a new kind of elementary particle. The identification of an indirect detection signal from DM in terms of cosmic radiation would surely be a fundamental breakthrough, with impact not only on astrophysics and cosmology but also on fundamental particle physics. 

From the observational side, the field is currently led by the \fermi, which has been operating for more than 10 years and has recently been prolonged in its operations. The excellent angular and energy resolutions of the telescope at gamma-ray energies, together with the capability to produce full-sky high-quality data, makes the \fermi\ the perfect detector to investigate the UGRB. Data from the \fermi\ have been used to determine the total intensity of the UGRB \cite{2015IGRB} and have been exploited to investigate its tiny fluctuations \cite{Ackermann:2012uf,Fornasa:2016ohl,Ackermann:2018wlo}.

In order to calculate the cross-correlation between the 21-cm brightness temperature and the gamma-ray intensity, we need to model the elements entering the power spectra and the gamma-ray window functions. We have two type of sources which are expected to contribute to the UGRB: $i)$ astrophysical sources, for which we will explicitly consider active galactic nuclei, further subdivided into BL Lac objects, misaligned active galactic nuclei (mAGN) and flat-spectrum radio quasars (FSRQ), plus star-forming galaxies (SFG); $ii)$ particle DM, for which we will consider the case of annihilating DM, a case relevant for WIMP. The window function for gamma rays from annihilating DM is (see e.g.\  Refs.~\cite{Fornengo2014,Camera:2014rja}):
 \begin{equation}
     W_{\gamma,\rm DM}(E, z) = \frac{1}{4 \pi} \, \frac{\langle \sigma v \rangle}{2} \Delta^2(z) \, \left (\frac{\Omega_{\rm DM} \,\rho_{\rm c} }{m_{\chi}} \right)^2 (1+z)^3  \frac{\de N}{\de E}[(1+z)E]  {\rm e}^{-\tau \left[(1+z)E,  z \right]},
     \label{eqn:Wann}
 \end{equation}
where $m_\chi$ and $\sigmav$ are the DM mass and annihilation cross-section, $\de N/\de E$ is the number of photons produced per annihilation event, $\Omega_{\rm DM}$ denotes the density of DM in the Universe today in units of the critical density $\rho_{\rm c}$, and $\tau$ is the optical depth, which takes into account the photon absorption for pair-production on the extragalactic background light, for which we adopt the parameterisation of Ref.~\cite{finke}. The clumping factor $\Delta^2(z)$ is introduced since the DM annihilation signal probes the DM density squared and $\langle\rho^2\rangle \ne \langle\rho\rangle^2$. It is defined as:
\begin{equation}
\Delta^2(z) \,\equiv \,\frac{\langle\rho^2\rangle}{\bar{\rho}^2}\, =\, \int_{{M_{\rm min}}}^{{M_{\rm max}}} \de M\, \frac{\de N}{\de M}(M,z)\int \de^3x \;\frac{\rho^2({\bm x}|M)}{\bar{\rho}^2},
\end{equation}
where $\rho({\bm x}|M)$ is the density profile of a DM halo of mass $M$ and $\bar\rho=\langle\rho\rangle$ is its average value. We describe the DM halo with a NFW profile \cite{NFW} with a concentration parameter related to the halo mass as derived in Ref.~\cite{correa} for the low-redshift regime ($z \le 4$). We also allow for the presence of substructures within main haloes by replacing $\rho^2 ({\bm x}|M)$ by $[1+B(M, z)] \rho^2 ({\bm x}|M, z)$ where $B(M, z)$ is the boost factor, for which we adopt the parameterisation of Ref.~\cite{boost}.

The window function for astrophysical sources is obtained from their differential flux $\de F/\de E$ weighted by their GLF $\phi(L,z)$, thus giving:

\begin{equation}
     W_{\gamma, \star}(E, z) = \left(\frac{d_L(z)}{1+z} \right)^2 \int_{L_{\rm min}}^{L_{\rm max}} \de L \, \frac{\de F}{\de E}(E,L,z) \, \phi(L,z), \label{eq:wastro}
\end{equation}
where $d_L = \chi(z) \times (1+z)$ is the luminosity distance. The minimum and maximum luminosities $L_{\rm min}$ and $L_{\rm max}$ in the integral depend on the intrinsic properties of the source class. However, $L_{\rm max}$ cannot be larger than the $L_{\rm sens}$ above which a source class can be resolved by the detector, in which case the upper bound is floored to $L_{\rm sens}$ (if also $L_{\rm min}$ is larger than $L_{\rm sens}$, the mean luminosity and the window function clearly vanish, since this would imply that all sources of that class are resolved). $L_{\rm sens}$ is determined for each class of sources in accordance with the detector flux sensitivity, for which we assume $F_{\rm sens} = 10^{-10}\,\mathrm{cm^2\,s^{-1}}$ for photons in the energy interval $1-100\,\mathrm{GeV}$, well compatible with the sensitivity of \fermi\ in 8 years of data taking. The GLFs, the spectral indices, and the minimum and maximum luminosities for the four types of astrophysical sources under consideration are reported in Table~\ref{tab:astroparam}. The mass-luminosity relations are taken from Ref.~\cite{Camera:2014rja}.
\begin{table}[t]
\centering
\begin{tabular}{lcccc}
\hline
~& $\alpha$ & $L_{\rm min}\,[\mathrm{erg\,s^{-1}}]$ & $L_{\rm max}\,[\mathrm{erg\,s^{-1}}]$) & Ref. \\
\hline
BL Lacs & 2.11 & $7 \times 10^{43}$ & $10^{52}$ & \cite{Lac} \\
FSRQ & 2.44 & $10^{44}$ & $10^{52}$ & \cite{FSRQ} \\
mAGN & 2.37 & $10^{40}$ & $10^{50}$ & \cite{mAGN1, mAGN2} \\
SFG & 2.7 & $10^{37}$ & $10^{42}$ & \cite{SFG} \\
\hline
\end{tabular}
\caption{Spectral index $\alpha$, minimum luminosity $L_{\rm min}$ and maximum luminosity $L_{\rm max}$ and reference for the gamma-ray luminosity function for the classes of astrophysical sources adopted in our analysis.}
\label{tab:astroparam}
\end{table}

\begin{figure}[t]
\centering
\includegraphics[width=0.49\textwidth]{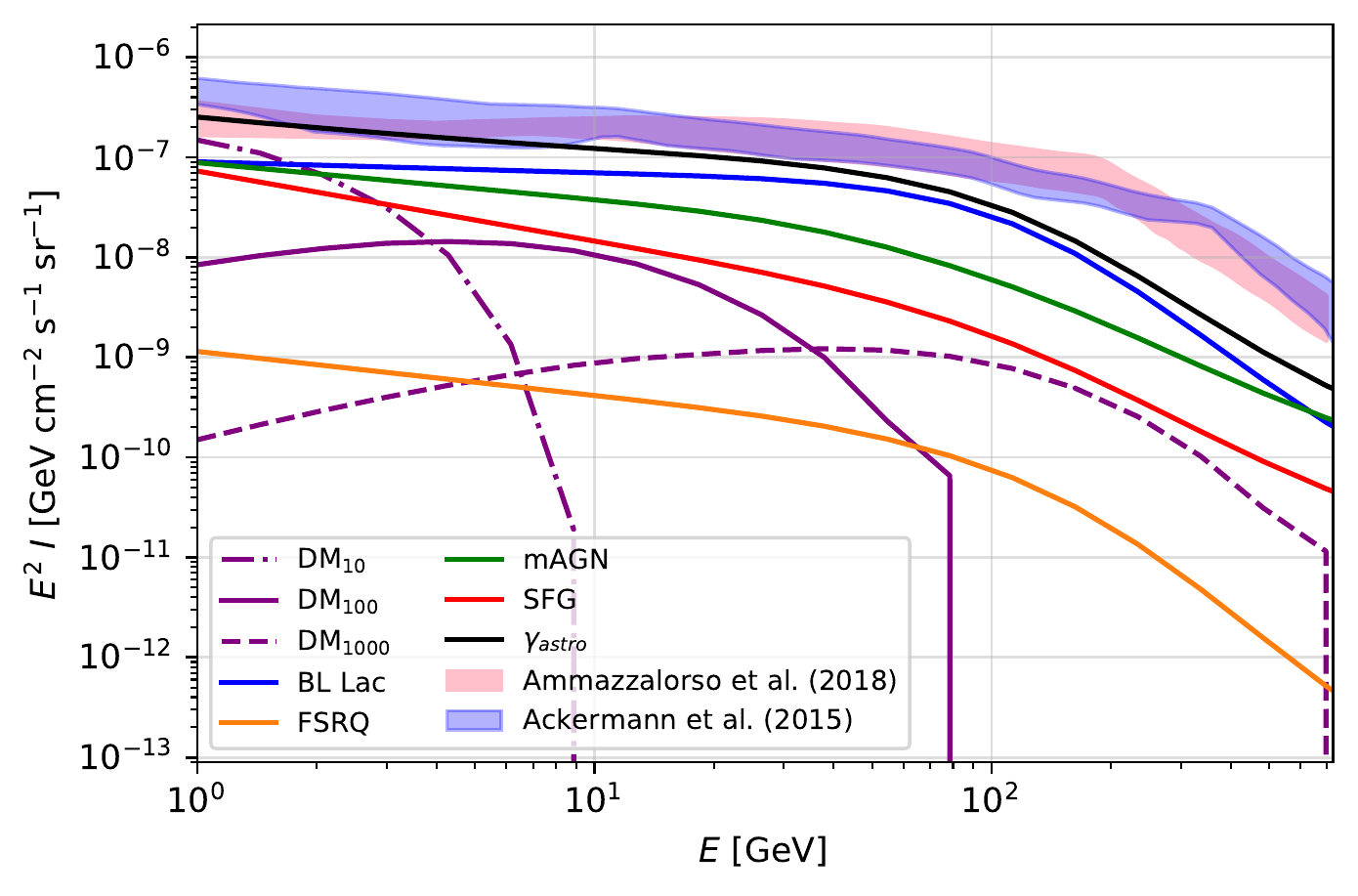}
\includegraphics[width=0.49\textwidth]{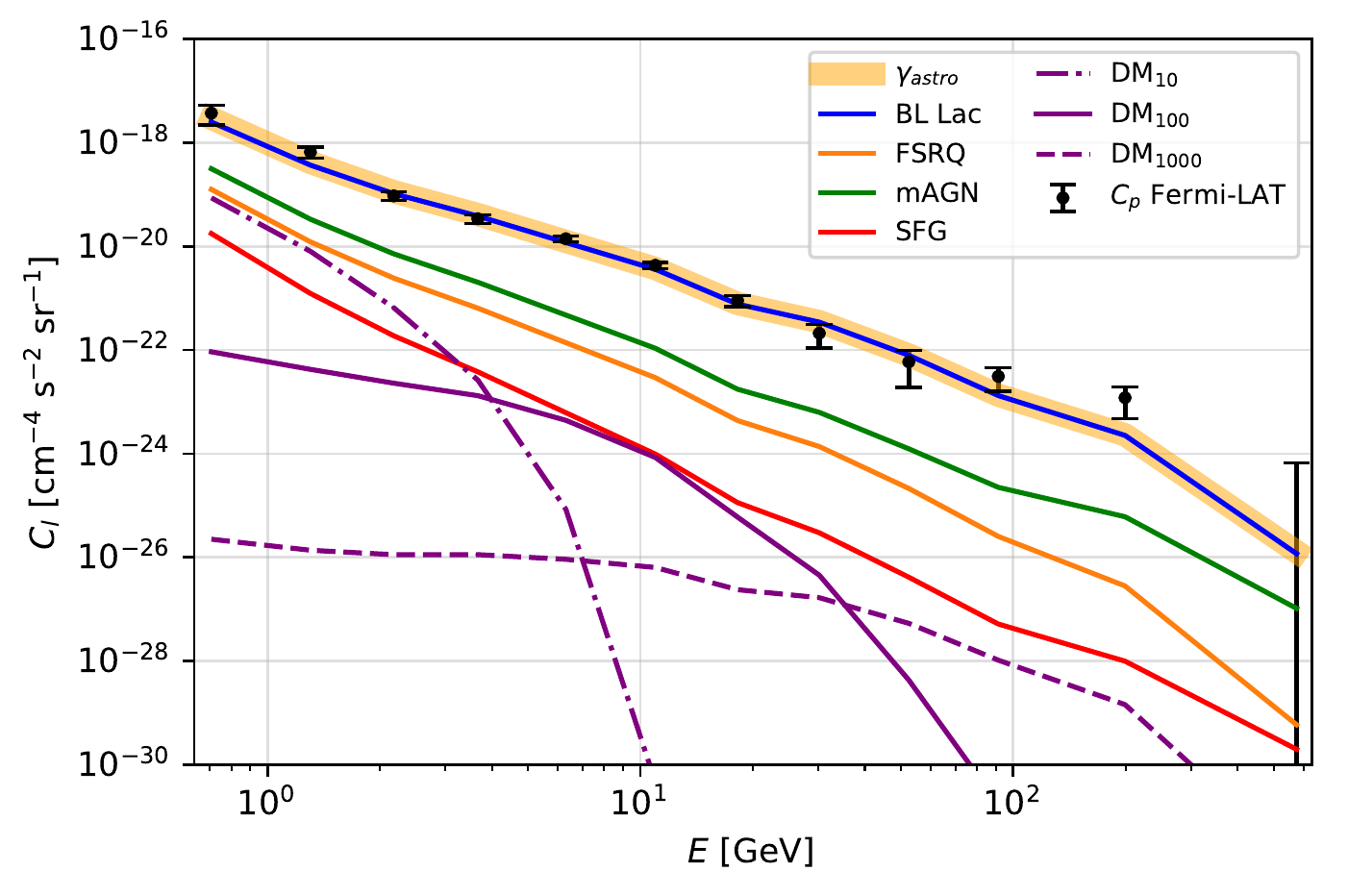}
\caption{Left panel: Measured total astrophysical flux (pink \cite{Ammazzalorso:2018evf} and blue \cite{2015IGRB} shaded bands) together with the theoretical predictions for the contribution from BL Lac (blue), FSRQ (orange), mAGN (green) and SFG (red) as a function of energy. The upper black line denotes the sum of all astrophysical components. The purple dot-dashed, solid and dashed lines show three predictions for a DM signal produced by particles with thermal cross-section and mass $m_{\chi}=10, 100, 1000$ GeV, respectively. 
Right panel: Comparison between the theoretical estimation for the auto-correlation APS $C_l^{\gamma \gamma}$ (black) and the latest measurement of the Fermi-LAT $C_{\rm P}$ \cite{Ackermann:2018wlo} (dots with error bars) as a function of the energy bin, together with the expected contributions from BL Lac (blue), FSRQ (orange), mAGN (green) and SFG (red). The orange thick line refers to the sum of all the astrophysical contributions, and almost coincides with the BL Lac case. The purple dot-dashed, solid and dashed lines show three predictions for a DM signal produced by particles with thermal cross-section and mass $m_{\chi}=10, 100, 1000$ GeV, respectively. The annihilation channel is $\bar b b$ through the paper.}
\label{fig:gamma}
\end{figure}

We validate the gamma-ray modelling by comparing in the left panel of Fig.~\ref{fig:gamma} the average UGRB intensity, $\langle I_\gamma \rangle = \int\de z \,c \, W_\gamma(z)/H(z)$, in the energy bins of Table~\ref{tab:noise_gamma} with the total intensity reported by the \fermi\ Collaboration \cite{2015IGRB} and the results obtained in Ref.~\cite{Ammazzalorso:2018evf} with a \fermi\ data set similar to the one adopted in our analysis. We find that our nominal model well reproduces the observations. The total intensity shown in the figure refers to the sum of the astrophysical components only, demonstrating that the model is able to reproduce the UGRB intensity without DM. The three DM cases are shown for comparison and refer to DM particles with $m_{\chi} = 10, 100, 1000$ GeV and thermal annihilation cross-section $\sigmav = 3\times 10^{-26}\,\mathrm{cm^3\,s^{-1}}$ in the $\bar bb$ channel. This will be the study-case adopted troughout the paper. They outline the different energy behaviours between the astrophysical sources and DM emissions, a fact that can be exploited also in the cross-correlation analysis to disentangle the two types of contributions. This also shows that the DM contribution to the UGRB mean intensity is expected to be subdominant.

In order to further check the models also at the level of produced anisotropies, we calculate the auto-correlation APS and compare them with the measurements of the anisotropies of the UGRB obtained in Ref.~\cite{Ackermann:2018wlo}. To calculate the gamma-ray auto-correlation APS, we need to specify the three-dimensional power spectra, whose ingredients will also be used in the Sec.~\ref{sec:results} to determine the cross-correlation signal with the 21-cm line.
The power spectra for annihilating DM are obtained from the Fourier transform of the matter density squared. In the halo model we again decompose the power spectrum into its 1-halo and 2-halo terms as follows:
\begin{align}
P^{\rm 1h}_{\rm DM}(k,z) &= \int_{M_{\rm min}}^{M_{\rm max}}\de M\, \frac{\de N}{\de M}(M,z) \,\left(\frac{\widetilde{v}_{\rm DM}(k|M,z)}{\Delta^2(z)} \right)^2,\\
P^{\rm 2h}_{\rm DM}(k,z) &= \left[\int_{M_{\rm min}}^{M_{\rm max}} \de M\, \frac{\de N}{\de M}(M,z) \, b(z,M) \,\left(\frac{\widetilde{v}_{\rm DM}(k|M,z)}{\Delta^2(z)} \right)\right]^2 \,P_{\rm lin}(k,z),
\end{align}
where $\widetilde{v}_{\rm DM}(k,M,z)$ is the Fourier transform of $[1+B(M,z)]\rho^2({\bm x}|M)/\overline{\rho}^2$ truncated at the virial radius $R_{\rm vir}$.

\begin{figure}[t]
\centering
\includegraphics[width=0.49\textwidth]{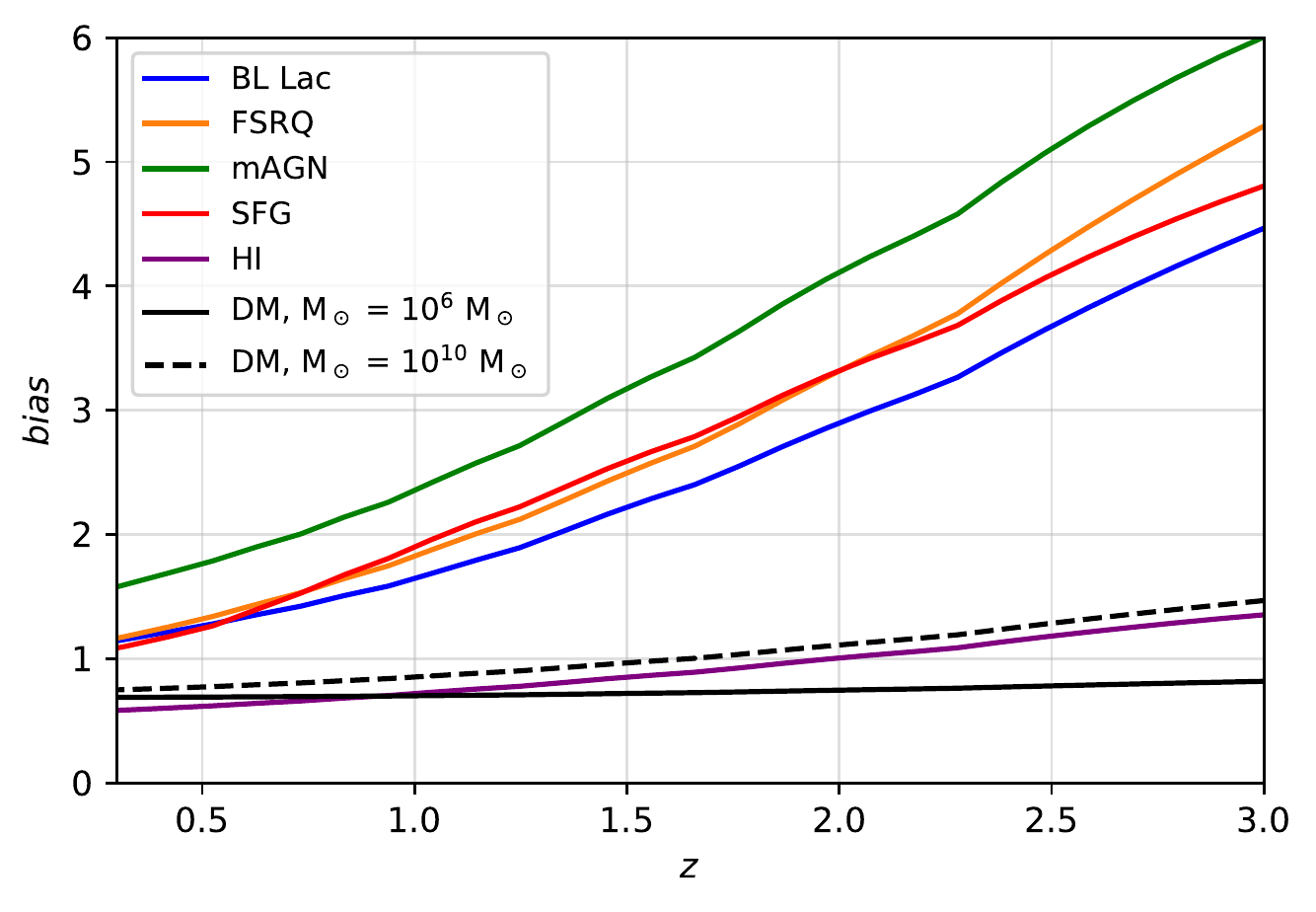}
\includegraphics[width=0.49\textwidth]{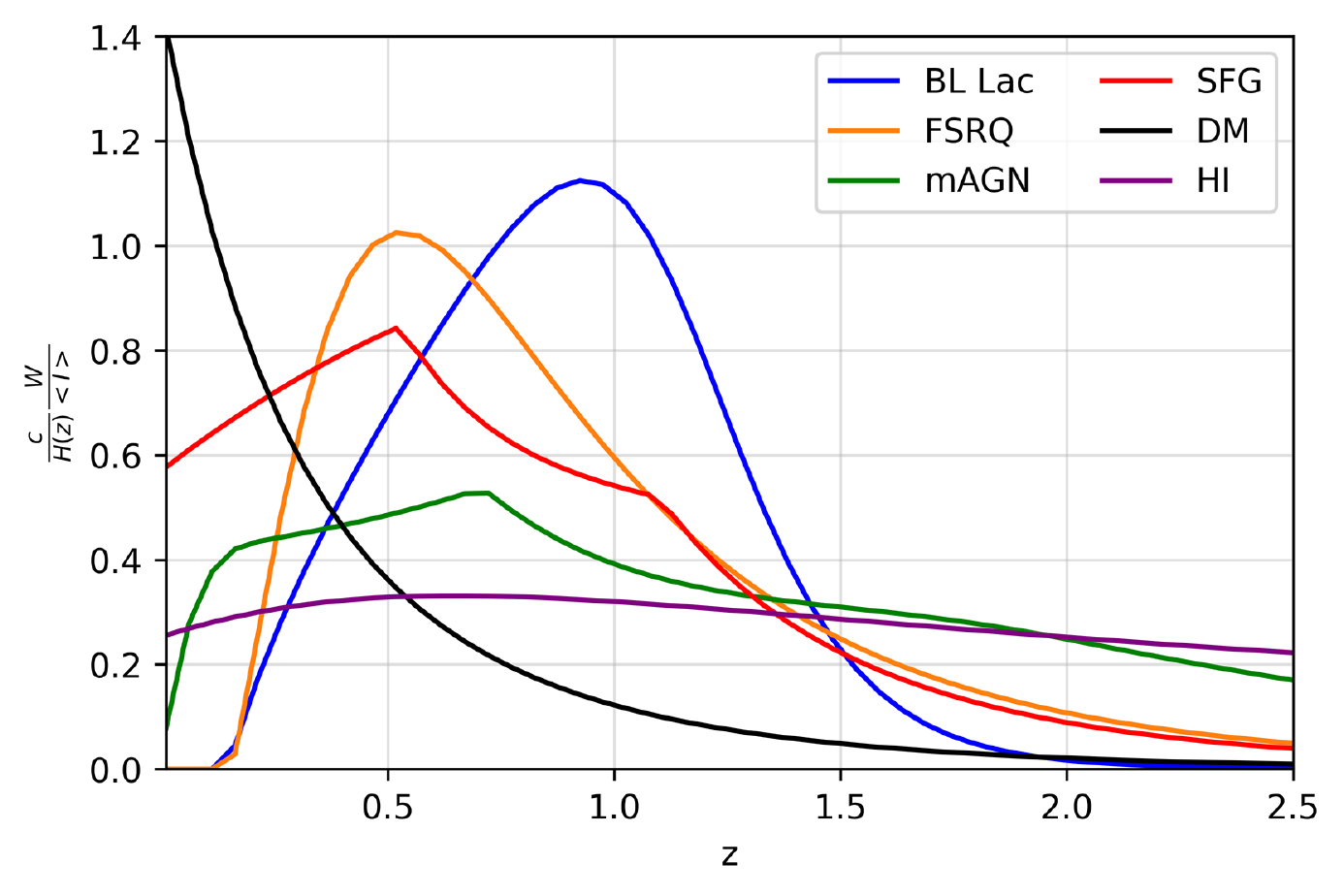}
\caption{Left panel: Bias as a function of redshift for BL Lac (blue), FSRQ (orange), mAGN (green), SFG (red), DM for haloes of mass $M = 10^{6}\,M_\odot$ (solid black) and $10^{10}\,M_\odot$ (dashed black).  Right panel: Normalized window functions as a function of redshift for the gamma-ray emitters, namely annihilating DM (black), BL Lac (blue), FSRQ (orange), mAGN (green), SFG (red), and \hi\ (purple). For gamma rays, the window function refers to $E=5$ GeV.} 
\label{fig:bias}
\end{figure}

For astrophysical sources, we have:
\begin{align}
P_\star^{\rm 1h}(k,z) &=  \int_{L_{\rm min}}^{L_{\rm max}} \de L  \,\phi(L,z) \left(\frac{L}{\langle g_\star\rangle} \right)^2,\\
P_\star^{\rm 2h}(k,z) &= \left[ \int_{L_{\rm min}}^{L_{\rm max}} \de L \,\phi(L,z) \, b(M(L),z) \,\frac{L}{\langle g_\star \rangle} \right]^2 P_{\rm lin}(k,z),
\end{align}
where $L_{\rm min}$ and $L_{\rm max}$ are subject to the consideration done in relation to Eq.~(\ref{eq:wastro}). The bias of astrophysical sources $b_\star$ can be defined as:
\begin{equation}
    b_\star(z) = \int_{L_{\rm min}}^{L_{\rm max}} \de L \, \phi(L,z) \,b(M(L),z) \,\frac{L}{\langle g_\star \rangle}
\end{equation}
and is shown in the left panel of Fig.~\ref{fig:bias}. 

For anisotropies, too, we validate the gamma-ray modelling by comparing in the right panel of Fig.~\ref{fig:gamma} the magnitude of anisotropy of the UGRB  $C^{\gamma}_{\rm P}(E)$ obtained by our models of astrophysical sources with the corresponding measurements from \fermi\ data \cite{Ackermann:2018wlo}. We see that a quite good agreement is obtained. The three DM cases are shown for comparison and refer to the same DM particle masses and thermal annihilation cross-section adopted in the left panel of Fig.~\ref{fig:gamma}. Again, they outline the different energy behaviour between the astrophysical sources and DM emissions, and show that the DM contribution to the UGRB auto-correlation is expected to be subdominant.

To conclude, we specify for which type of gamma-ray detector we will derive our forecasts for the cross-correlation. Our reference will be the \fermi\ specifications adopted for the anisotropy analysis of the UGRB performed in Ref.~\cite{Ackermann:2018wlo}. They are based on 8 years of data taking and a selection of events with optimal angular resolution and background rejection. We refer to Ref.~\cite{Ackermann:2018wlo} for a thorough display of all specifications. We determine our cross-correlation predictions on the 12 energy bins reported in Table~\ref{tab:noise_gamma}. For each energy bin, Table~\ref{tab:noise_gamma} also shows the photon noise $N^\gamma$, relevant to determine the error on the auto-correlation analysis, the sky fraction $f_{\rm sky}$ and the angular resolution of the \fermi\ detector $\sigma_0^{\rm Fermi}$ at the geometric centre of each bin $E = (E_{\rm min} E_{\rm max})^{1/2}$, as  determined by the \fermi\ Collaboration \footnote{https://www.slac.stanford.edu/exp/glast/groups/canda/lat\_Performance.htm}. The full \fermi\ beam window function depends on the photon event class and on the photon energy and is available through the \fermi\ tools. We found that a good analytic approximation for $B^\gamma_l(E)$ is a modified Gaussian in multipole-space, i.e.\ :
\begin{equation}
B^\gamma_l(E) = \exp \left[- \frac{\sigma_{\rm b}(l,E)^2 \, l^2}{2} \right],
\label{eq:beam1}
\end{equation}
where the dispersion evolves with $l$ for large multipoles as:
\begin{equation}
\sigma_{\rm b}(l, E) = \sigma_0^{\rm Fermi}(E) \left[1 + 0.25\,\sigma_0^{\rm Fermi}(E)\,l\right]^{-1}.
\label{eq:beam2}
\end{equation}
The normalisation $\sigma_0(E)$ is the 68\% containment angle of the \fermi\ detector at energy $E$. The energy evolution of the $\sigma_0(E)$ parameter is well reproduced by:
\begin{equation}
\sigma_0^{\rm Fermi}(E) = \sigma_0^{\rm Fermi}(E_{\rm ref})  \times (E/E_{\rm ref})^{-0.95} + 0.05\, {\rm deg},
\label{eq:beam3}
\end{equation}
with $E_{\rm ref}= 0.5$ GeV and $\sigma_0^{\rm Fermi}(E_{\rm ref}) = 1.20$ deg. The relation follows a power law behaviour and flattens out at around 0.05 deg for high energies (we adhered to the specifications of the PSF2 type response function, for definiteness). These empirical relations well reproduce the beam function adopted in Ref.~\cite{Ackermann:2018wlo} and derived from the \fermi\ data.

\section{Results}
\label{sec:results}
\begin{figure}[t]
\centering
\includegraphics[width=0.49\textwidth]{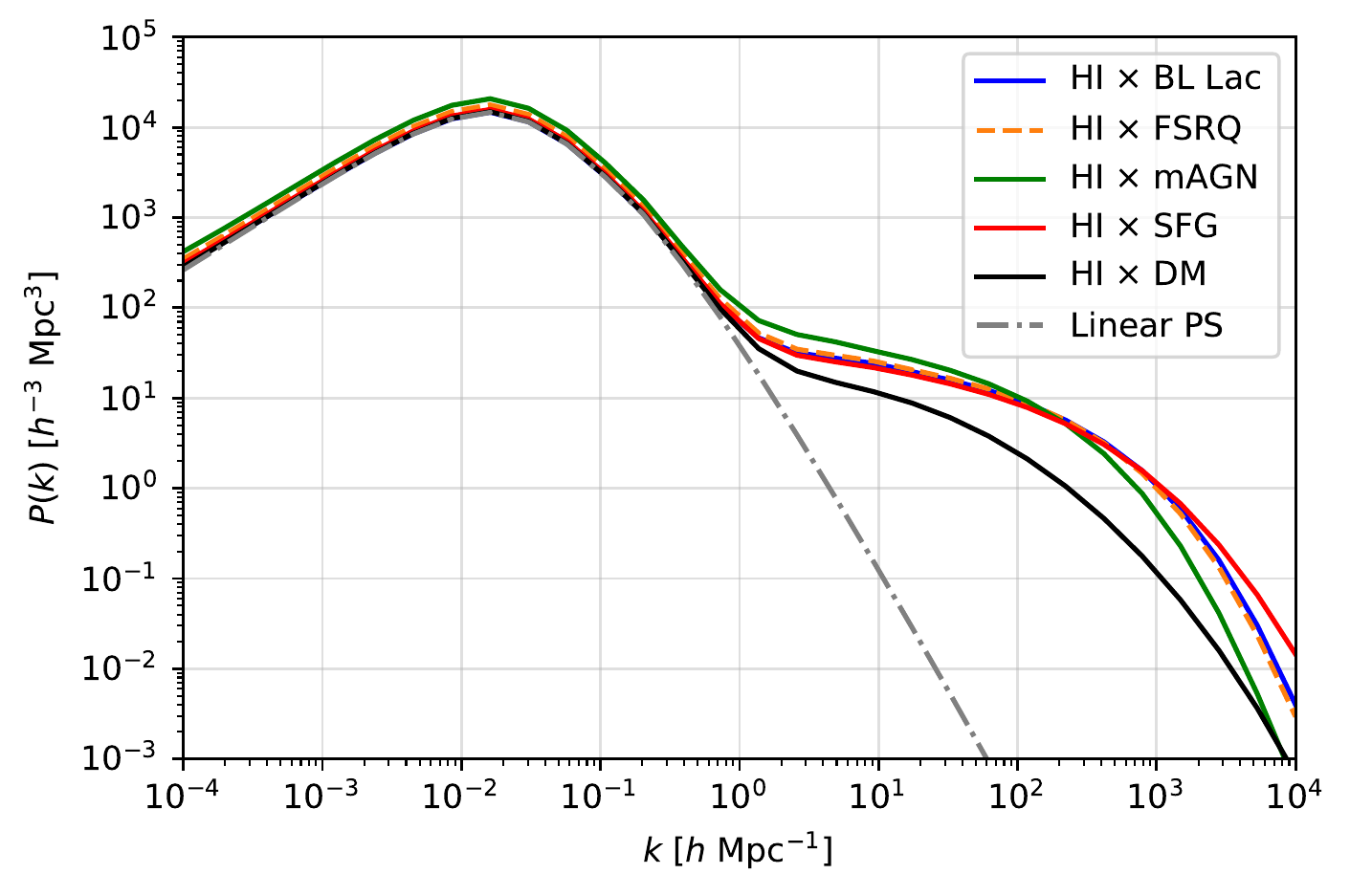}
\caption{Cross-correlation power spectra between the 21-cm-line brightness temperature and the different gamma-ray intensities emitted by annihilating DM (black line), BL Lac (blue), FSRQ (orange),  mAGN (green), SFG (red), calculated at $z=0.5$.}
\label{fig:PS_cross}
\end{figure}

Let us now turn to the discussion of the cross-correlation signal between the 21-cm-line  brightness temperature and the gamma-ray intensity. The power spectrum is built from the elements introduced in the previous sections for the two observables under discussion. For the cross-correlation with the DM gamma-ray emission, the expression is:
\begin{align}
    P^{\rm 1h}_{\rm HI-DM}(k,z) &=  \int_{M_{\rm min}}^{M_{\rm max}} \de M \, \frac{\de N}{\de M}(M,z) \frac{\tilde{v}_{\rm DM}(k|M,z)}{\Delta^2(z)} \tilde{u}_{\rm HI}(k|M,z) \frac{M_{\rm HI}(M(L),z)}{\overline{\rho}_{\rm HI}(z)},\\
    P^{\rm 2h}_{\rm HI-DM}(k,z) &=  \left[\int_{M_{\rm min}}^{M_{\rm max}}  \de M\, \frac{\de N}{\de M}(M,z) \frac{\tilde{v}_{\rm DM}(k|M,z)}{\Delta^2(z)}  b(z,M) \right]\nonumber\\
    &\times\left[\int_{M_{\rm min}}^{M_{\rm max}}  \de M \, \frac{\de N}{\de M}(M,z) \, \tilde{u}_{\rm HI}(k|M,z) \frac{M_{\rm HI}(M,z)}{\overline{\rho}_{\rm HI}} b(z,M) \right] P_{\rm lin}(k,z),
\end{align}
while for gamma-ray emission it becomes:
\begin{align}
    P^{\rm 1h}_{\rm HI\star}(k,z) &= \int_{L_{\rm min}}^{L_{\rm max}} \de L \, \frac{L}{\langle g_\star\rangle} \, \tilde{u}_{\rm HI}(M(L),z(  \phi(L,z) \,\frac{M_{\rm HI} (M(L),z)}{\overline{\rho}_{\rm HI}},\\
    P^{\rm 2h}_{\rm HI\star}(k,z) &= \left[\int_{M_{\rm min}}^{M_{\rm max}} \de M \, \frac{\de N}{\de M}(M,z) \, \tilde{u}_{\rm HI}(k|M,z) \, \frac{M_{\rm HI}(M,z)}{\overline{\rho}_{\rm HI}(z)} \, b(z,M) \right] \nonumber \\
    &\times  \left[\int_{L_{\rm min}}^{L_{\rm max}} \de L \, \frac{L}{\langle g_\star \rangle} \, \phi(L,z) \, b(z,M(L)) \right] P_{\rm lin}(k,z).
\end{align}
The power spectra for all terms are shown in Fig.~\ref{fig:PS_cross} at $z=0.5$. At large physical scales (small $k$) the power spectra are dominated by the 2-halo term, which closely follows the linear matter power spectrum; at physical scales smaller than $1\,h^{-1}\,\mathrm{Mpc}$ (large $k$) the 1-halo term becomes dominant. 

The window functions are those discussed in Sec.~\ref{sec:hydrogen} for \hi\ and Sec.~\ref{sec:gamma} for gamma rays and displayed in the right panel of Fig.~\ref{fig:bias}. We see that gamma-ray window functions of unresolved astrophysical sources peak at redshift in the range between 0.5 and 1, while for DM, whose emission is totally unresolved,  the peak is prominent at low redshift and quickly decays as the redshift increases. For the 21-cm brightness temperature, the window function is broad and almost featureless. However, the excellent frequency resolution of a radio telescope can be exploited to set apart specific redshift intervals: since the APS depends on the overlap of the window functions of the two observables, redshift tomography can then be used to outline the redshift range where DM or astrophysical emission is more prominent.
\begin{figure}[t]
\centering
\includegraphics[width=0.49\textwidth]{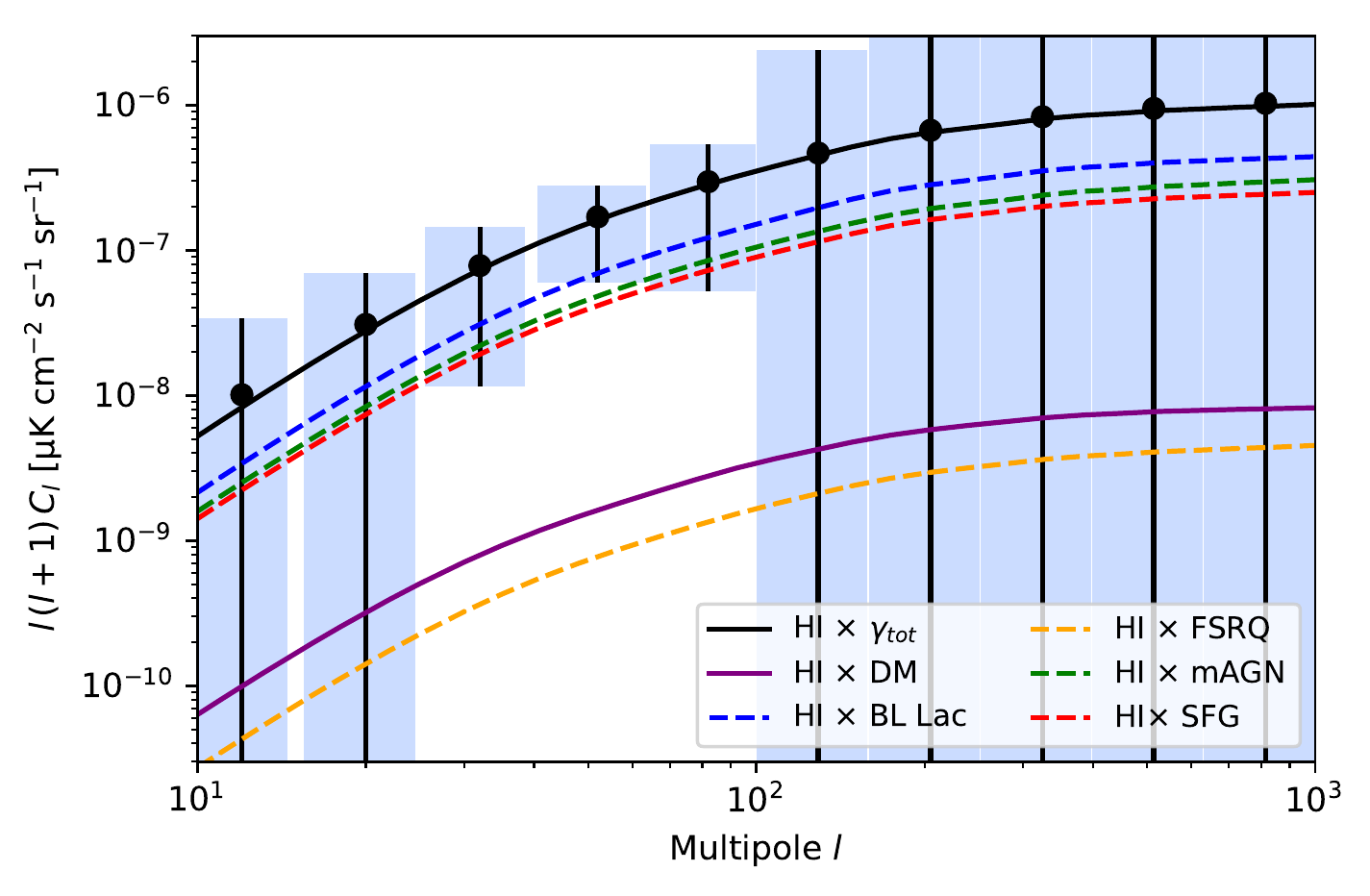}
\includegraphics[width=0.49\textwidth]{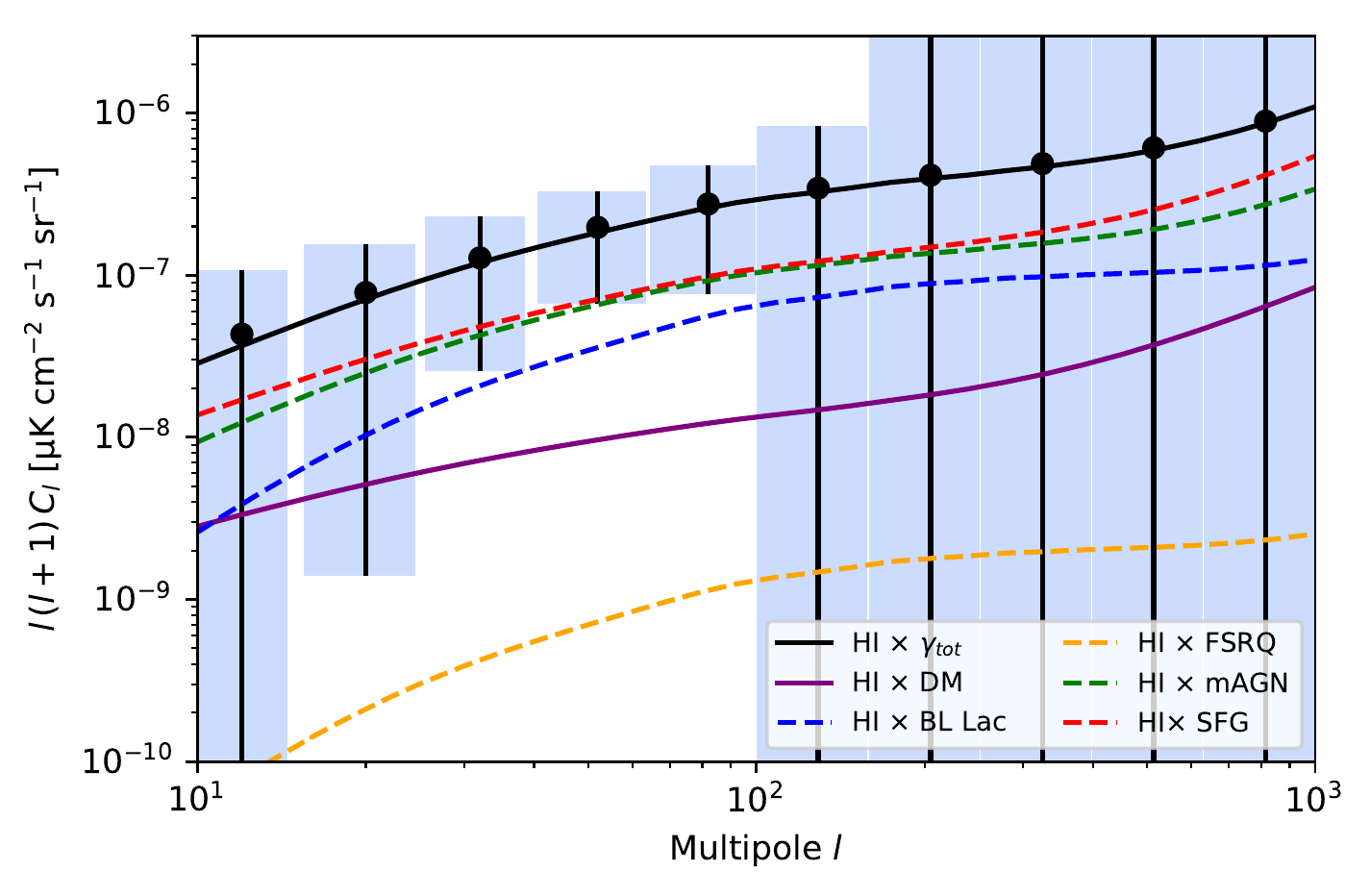}
\caption{Angular power spectrum of the cross-correlation between \hi\ intensity mapping and gamma rays. The different lines refer to the theoretical prediction of the signal originated by the different gamma rays sources, as indicated in the inset box. The purple solid line, which refers to the signal due to DM gamma-ray emission, is obtained for a DM mass $m_\chi = 100$ GeV and a thermal annihilation rate $\sigmav = 3\times 10^{-26}\,\mathrm{cm^3\,s^{-1}}$. The solid black line is the sum of all components. Error bars are obtained as the Gaussian estimate of the variance of the signal and refer to the combined (dish + interferometer) configuration. The results refer to the sum of the contributions of all \fermi\ energy bins of Table~\ref{tab:noise_gamma}. The radio telescope configuration is the higher redshift MeerKAT UHF-band in the left panel and the lower redshift  L-band in the right panel (bands are reported in Table~\ref{tab:specifiche_IM}).}
\label{fig:cross-meerkat}
\end{figure}

\begin{figure}[t]
\centering
\includegraphics[width=0.49\textwidth]{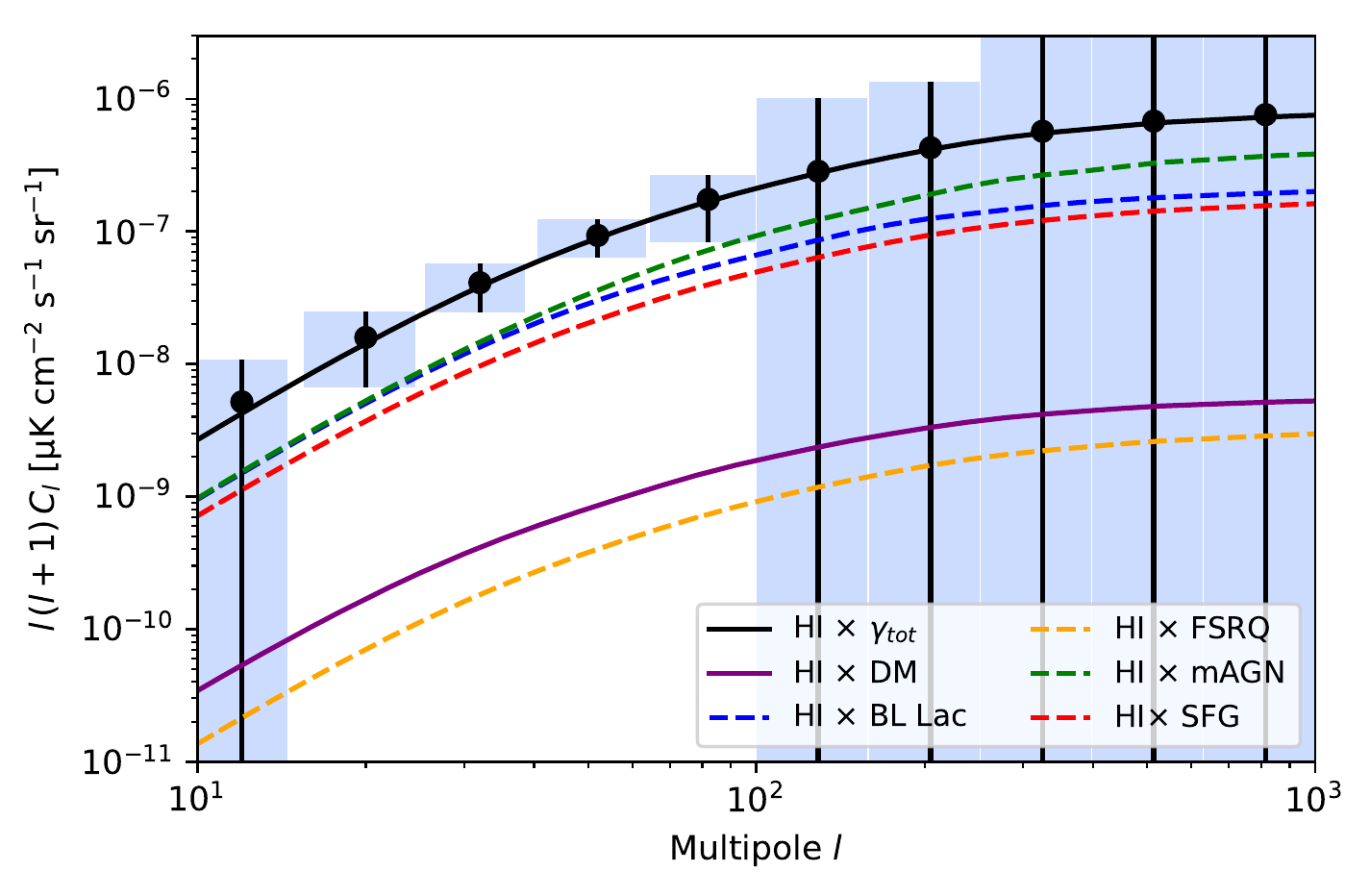}
\includegraphics[width=0.49\textwidth]{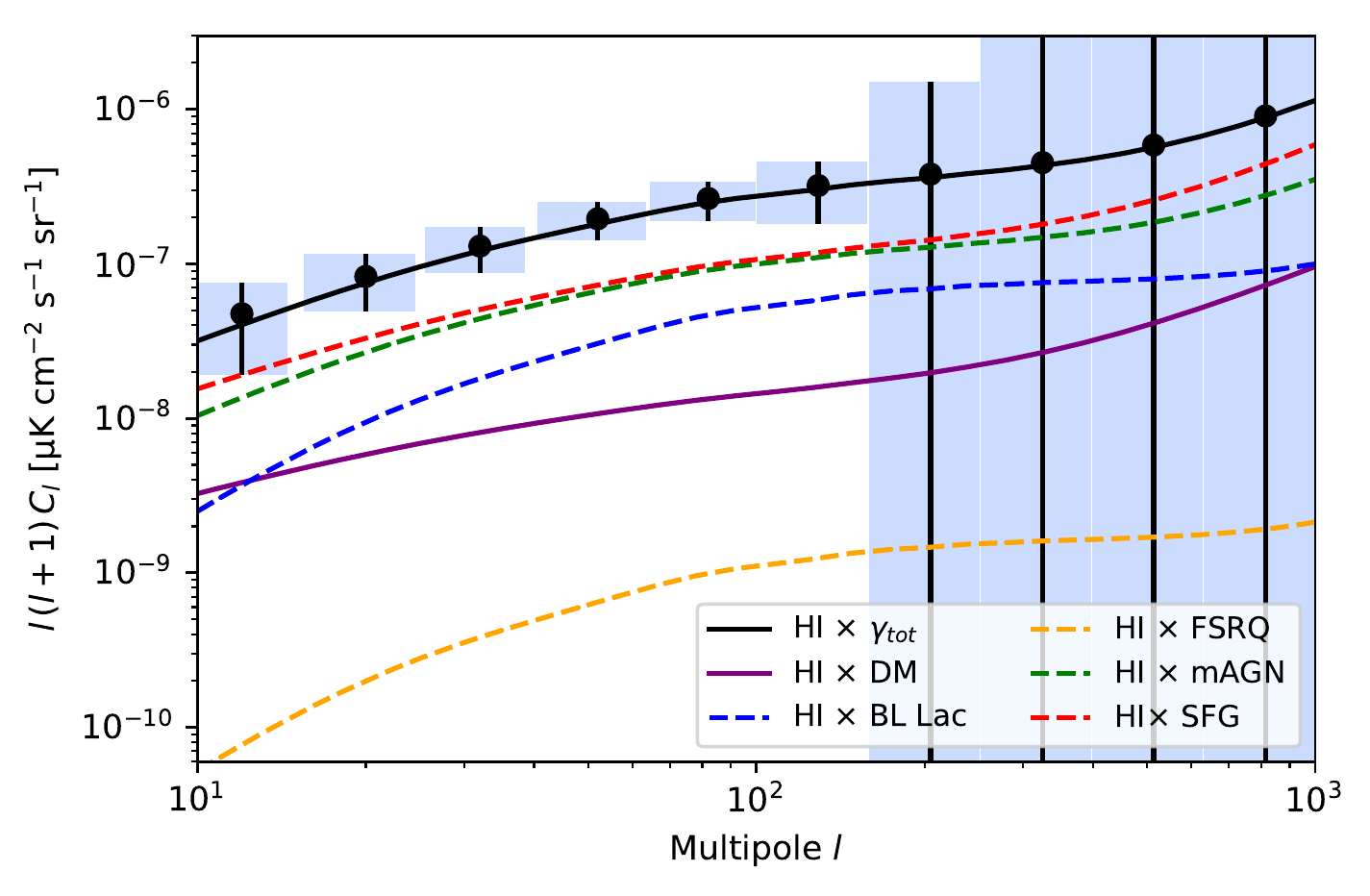}
\caption{The same as in Fig.~\ref{fig:cross-meerkat}, for SKA1 $\times$ {\it Fermi}-LAT. The left panel refers to the higher redshift SKA1 Band 1, the right panel to the lower redshift SKA1 Band 2.}
\label{fig:cross-ska1}
\end{figure}

\begin{figure}[t]
\centering
\includegraphics[width=0.49\textwidth]{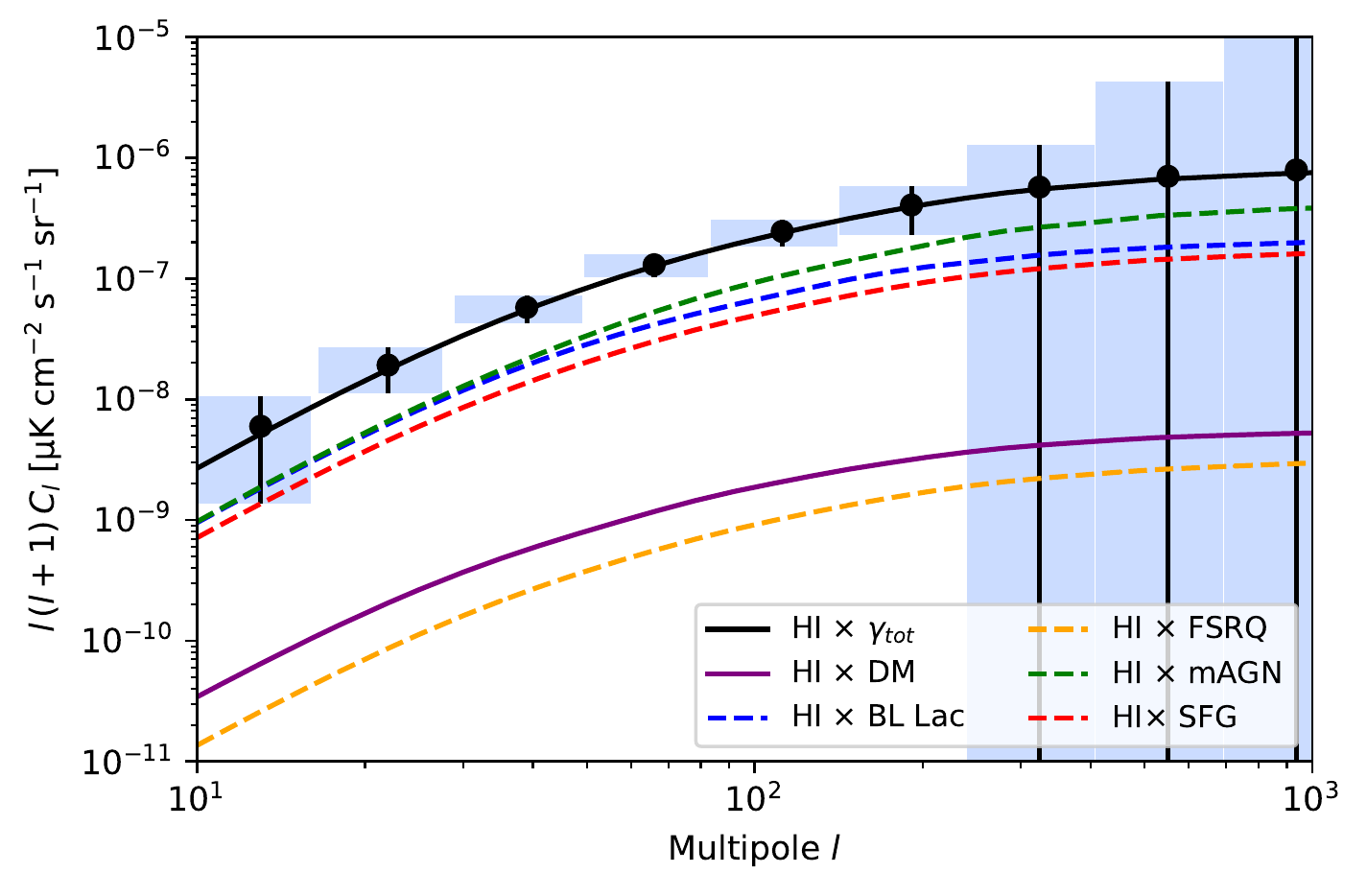}
\includegraphics[width=0.49\textwidth]{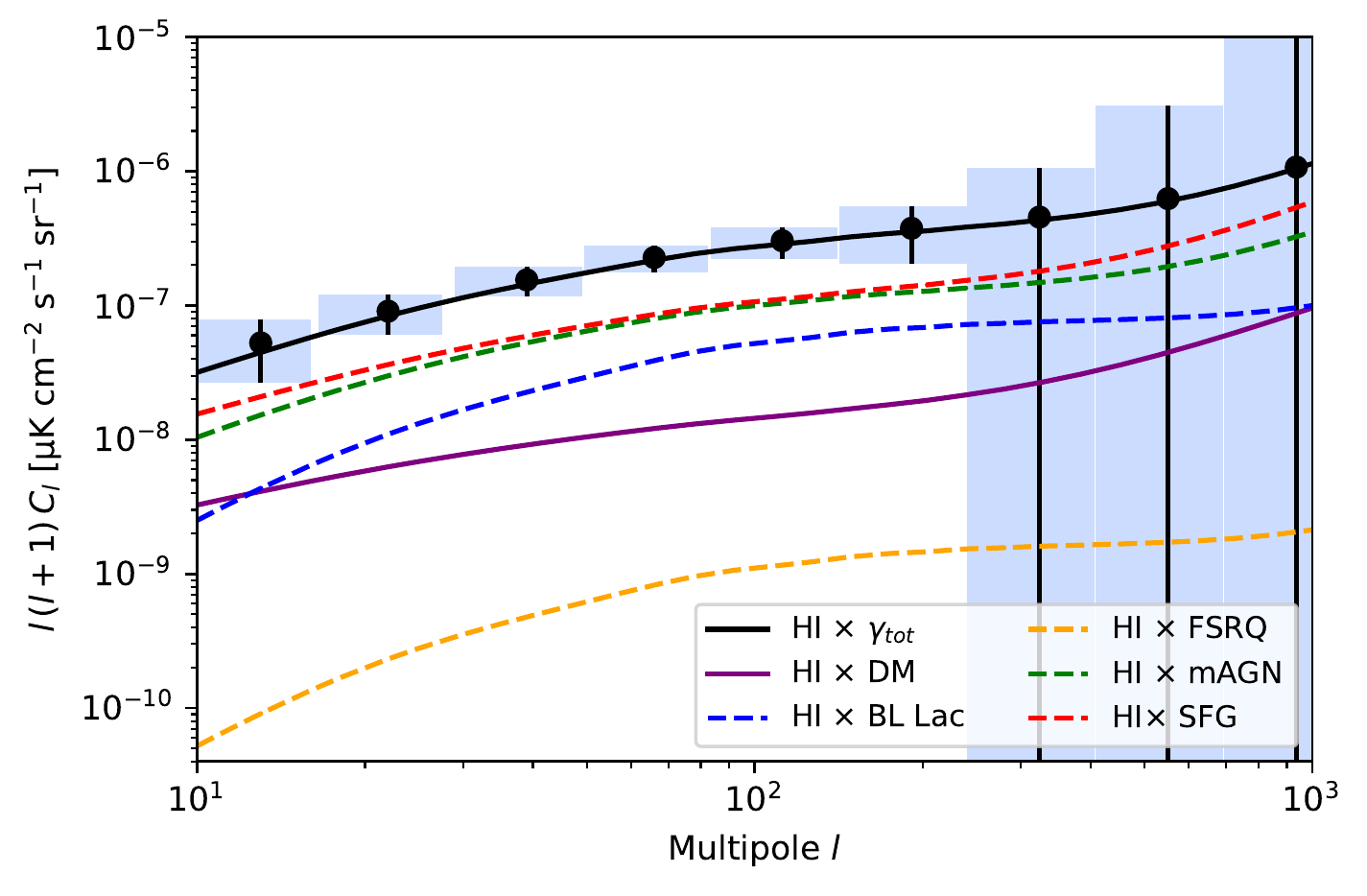}
\caption{The same as in Fig.~\ref{fig:cross-meerkat}, for SKA2 $\times$ {\it Fermi}-LAT. The left panel refers to the higher redshift SKA1 Band 1, the right panel to the lower redshift SKA1 Band 2.}
\label{fig:cross-ska2}
\end{figure}

The cross-correlation APS are shown in Fig.~\ref{fig:cross-meerkat} for the combination of \fermi\ with MeerKAT, and in Figs.~\ref{fig:cross-ska1} and \ref{fig:cross-ska2} for SKA1 and SKA2, respectively. In each figure, the left panel refers to the signal integrated in the higher redshift band (UHF-band for MeerKAT and Band 1 for SKA) and the right panel to the lower redshift band (L-band for MeerKAT and Band 2 for SKA), as reported in Table~\ref{tab:specifiche_IM}. The curves refer to the sum of the signal in the 12 gamma-ray energy bins of Table~\ref{tab:noise_gamma}. The dashed lines show the signal originated by astrophysical sources (as indicated in the inset boxes of the figures) and the (purple) solid line stands for the signal produced by the annihilation of a DM particle with mass $m_\chi = 100$ GeV and thermal cross-section $\langle \sigma v \rangle = 3 \times 10^{-26}\,\mathrm{cm^3\,s^{-1}}$ annihilating into a $b \bar b$ quark pair,  representative of an hadronic final state. The DM signal, being directly proportional to the annihilation cross-section, can be increased or decreased by acting on $\sigmav$, while a change in the mass values implies a different energy spectrum, as can be seen in Fig.~\ref{fig:gamma} for some representative cases. The DM clustering properties can also boost (by up to a factor of a few \cite{Gao:2011rf}) or suppress (by up to a factor of 10 \cite{Hiroshima:2018kfv}) the size of the APS of annihilating DM. 

\begin{figure}[t]
\centering
\includegraphics[width=0.49\textwidth]{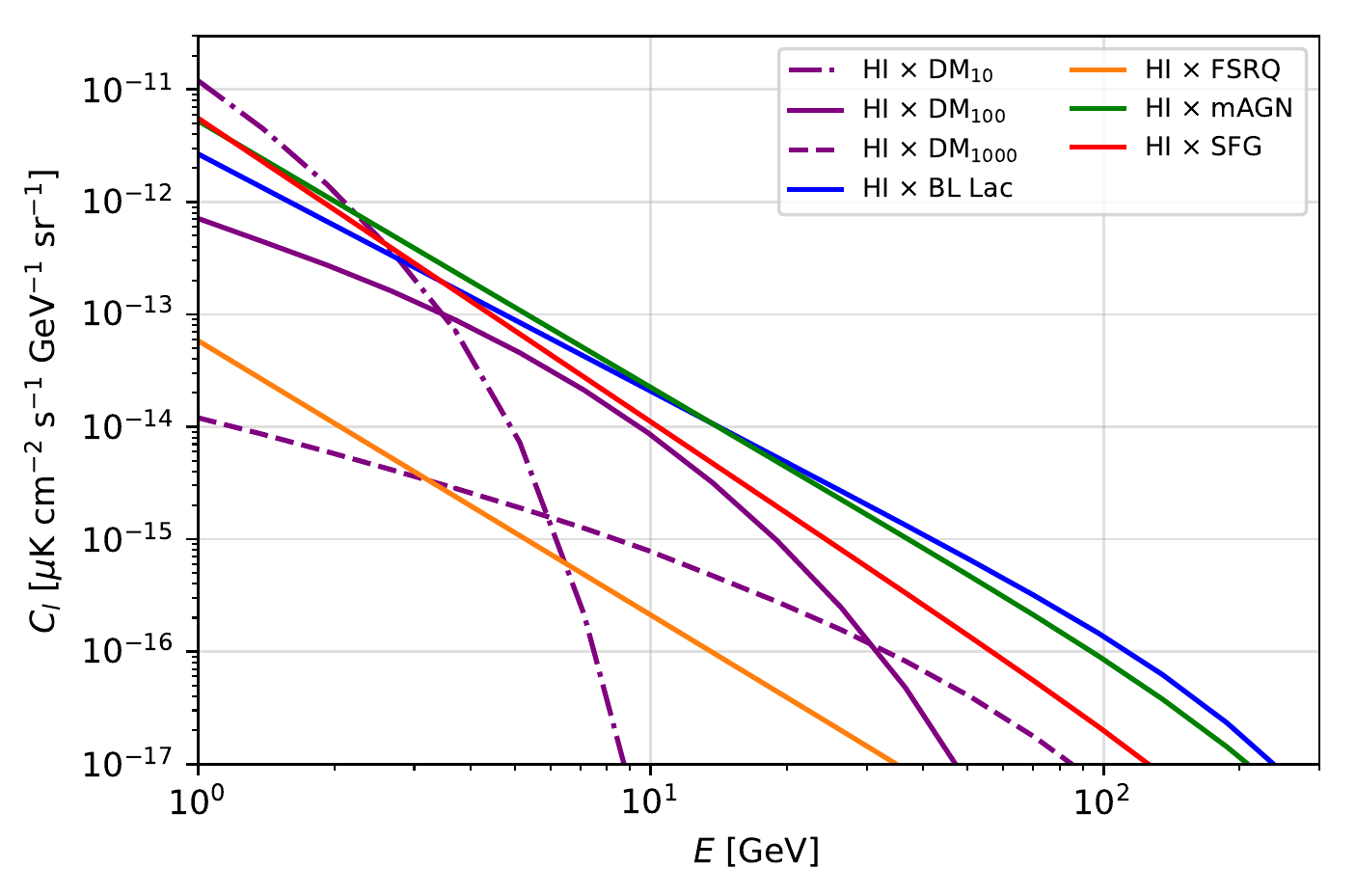}
\includegraphics[width=0.49\textwidth]{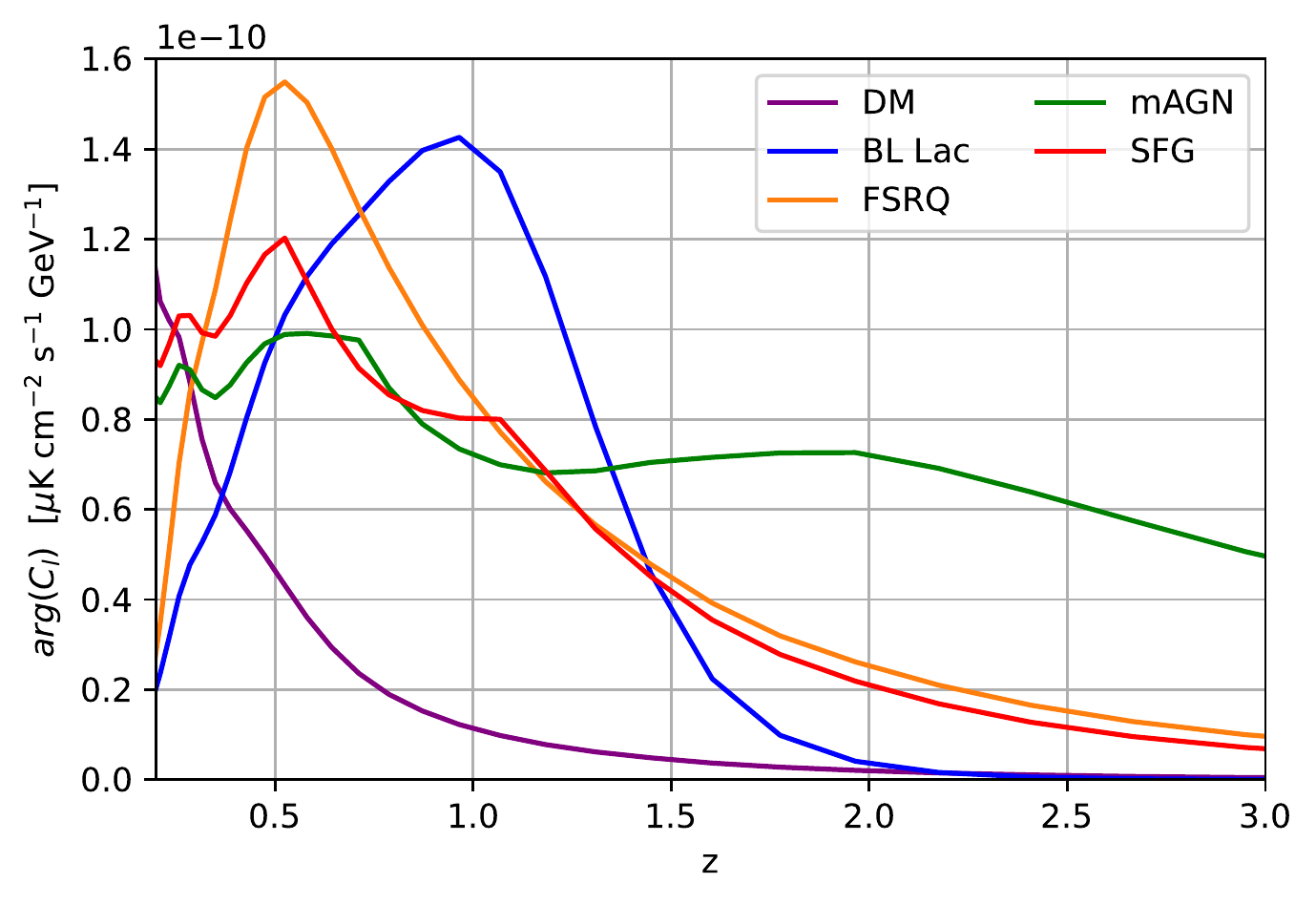}
\caption{Left panel: Energy behaviour of the angular power spectrum of the cross-correlation between \hi\ intensity mapping and gamma rays, for multipole $l=100$ and the redshift interval of SKA1 Band 2. Right panel: redshift dependence of the cross-correlation between \hi\ intensity mapping and gamma rays, for multipole $l=100$ and for photons with gamma-ray energy $E = 5$ GeV (and normalized to the gamma-ray and intensity mapping mean intensities $\langle I_\gamma \rangle  \langle I_{\hi} \rangle$, in order to make more visibile the relative behaviours) . For both panels the different lines refer to BL Lac (blue), FSRQ (orange), mAGN (green)  and SFG (red) as a function of energy.  The purple dot-dashed, solid and dashed lines in the left panel refer to a DM signal produced by particles with thermal cross-section and mass $m_{\chi}=10, 100, 1000$ GeV. In the right panel only the case $m_{\chi} = 100$ GeV is shown.}
\label{fig:Cl-behaviours}
\end{figure}

\begin{figure}[t]
\centering
\includegraphics[width=0.49\textwidth]{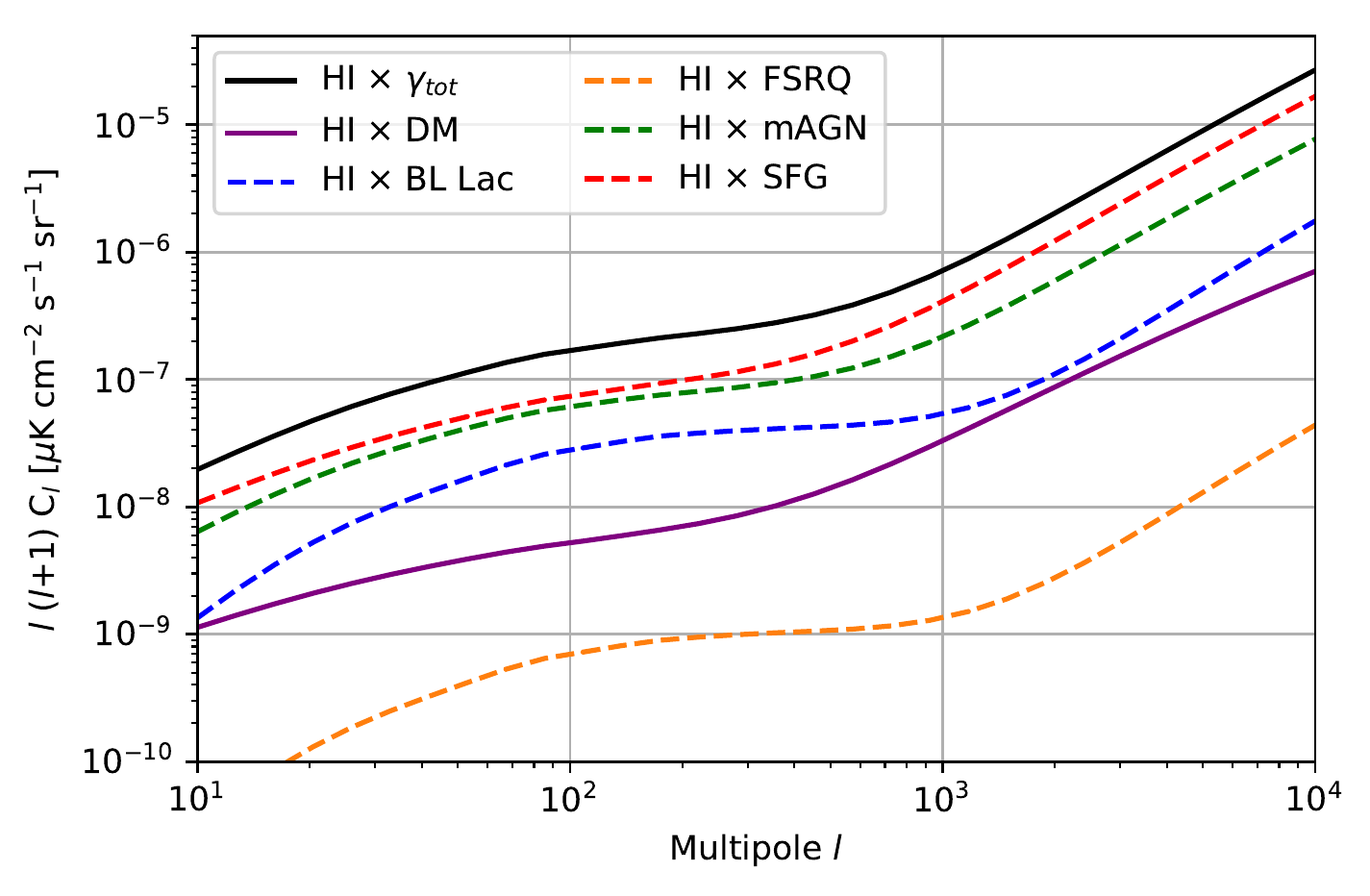}
\includegraphics[width=0.49\textwidth]{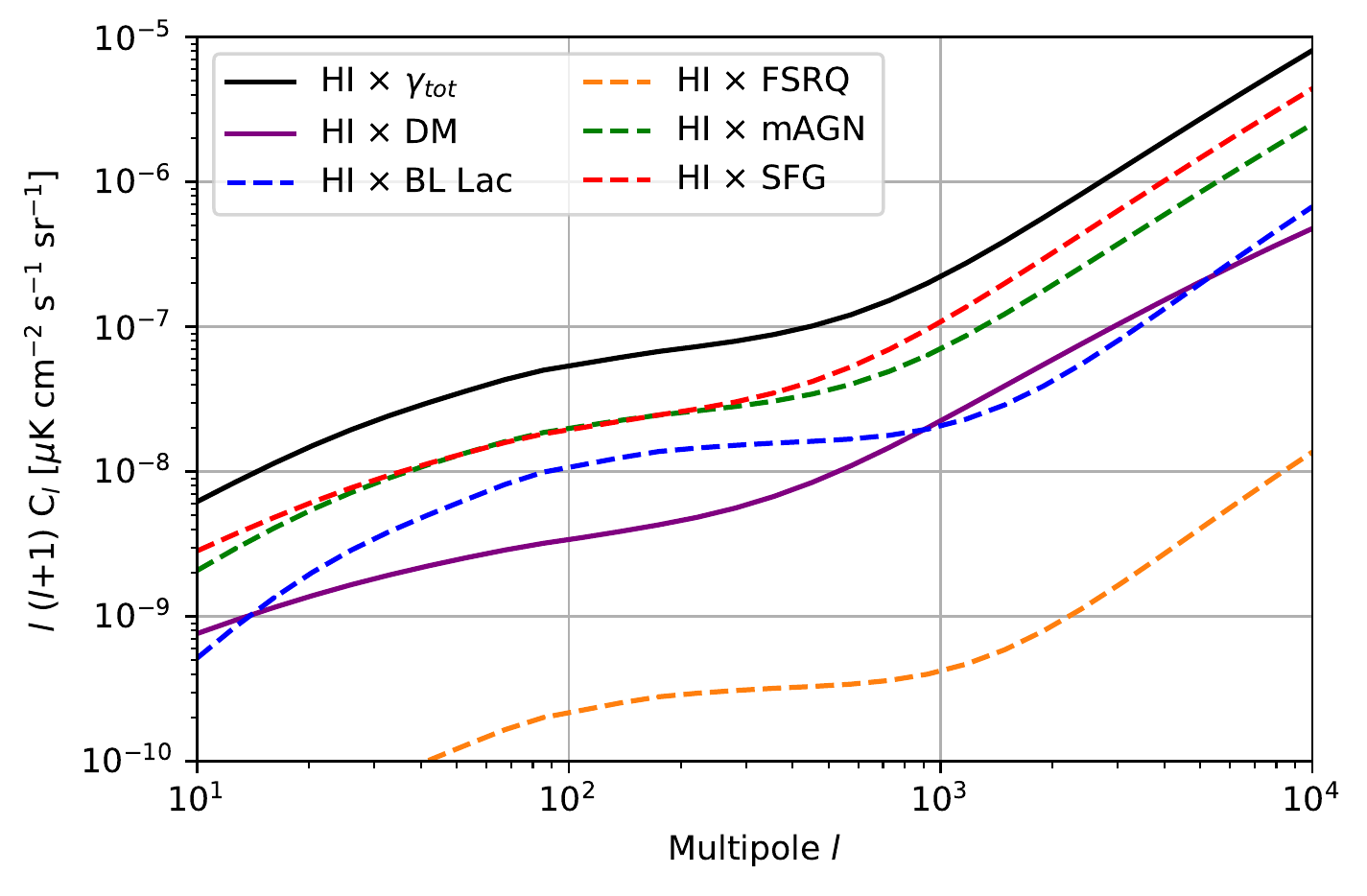}
\includegraphics[width=0.49\textwidth]{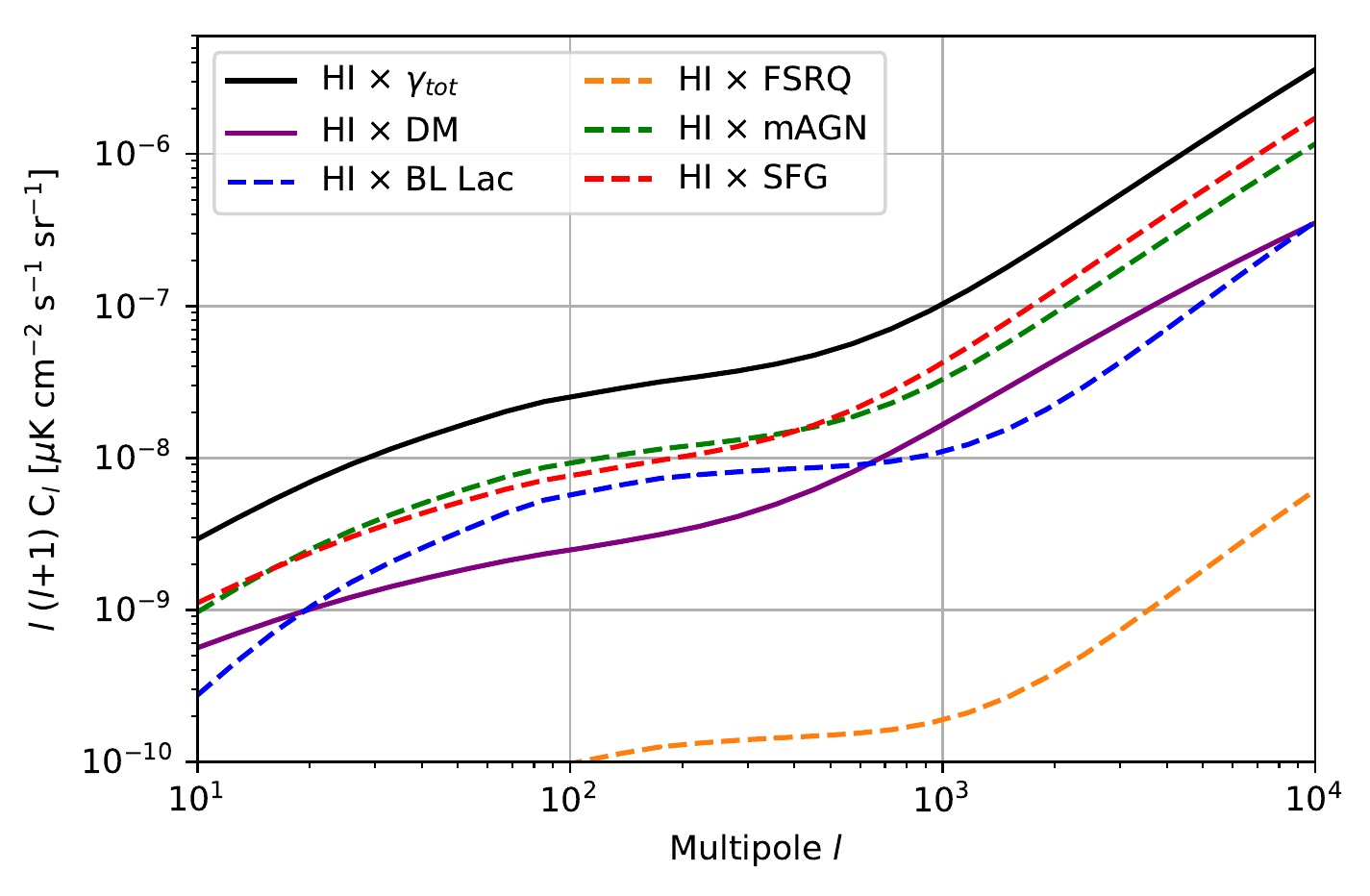}
\includegraphics[width=0.49\textwidth]{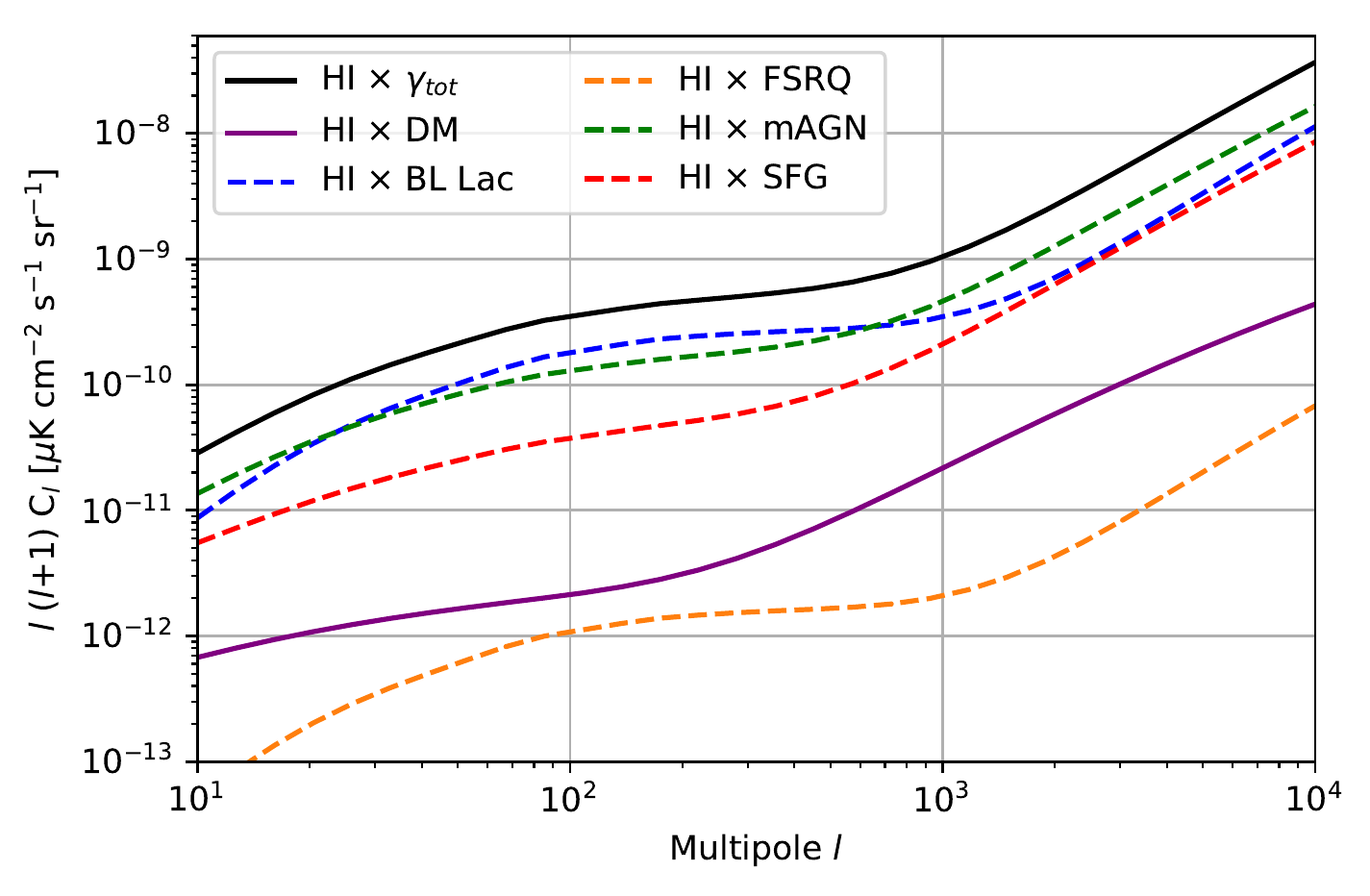}
\caption{Angular power spectrum of the cross-correlation between \hi\ intensity mapping and gamma rays. The different lines refer to the theoretical prediction of the signal originated by the different gamma rays sources, as indicated in the inset box. The purple solid line refers to the signal due to DM gamma-ray emission and is obtained for a DM mass $m_\chi = 100$ GeV and a thermal annihilation rate $\sigmav = 3\times 10^{-26}\,\mathrm{cm^3\,s^{-1}}$. The solid black line is the sum of all components. The results refer to the \fermi\ energy bins number 1, 2, 3 and 9 of Table~\ref{tab:noise_gamma}. The radio telescope configuration is the lower redshift Band 2 of SKA1, as reported in Table~\ref{tab:specifiche_IM}. }
\label{fig:ska1-breakdown}
\end{figure}

The relative size between the astrophysical sources and the DM signals in Figs. \ref{fig:cross-meerkat}, \ref{fig:cross-ska1} and \ref{fig:cross-ska2} are dictated by a complex interplay between the energy and redshift dependences of the angular power spectra $C_l$. In Fig. \ref{fig:Cl-behaviours} we show two sections of the 3-dimensional space $(l,E,z)$: the left panel shows the energy dependence of the angular power spectrum for multipole $l=100$ and the redshift interval of SKA1 Band 2; the right panel shows the redshift dependence of the angular power spectrum again for $l=100$ and for photon energy $E = 5$ GeV. In the right panel, the $C_l$ have been normalised to the product of the gamma-ray and intensity mapping mean intensities $\langle I_\gamma \rangle  \langle I_{\hi} \rangle$, in order to make more visibile the relative behaviours. A breakdown of the $C_l$ signal produced in different energy bins is instead shown in Fig.~\ref{fig:ska1-breakdown} for the lower redshift band Band 2, which shows the various contributions at different energies. Concerning the redshift dependence, the comparison of the left and right panels of Figs.~\ref{fig:cross-meerkat}, \ref{fig:cross-ska1} and \ref{fig:cross-ska2} shows that the signal coming from higher redshift (Band 1) is dominated by the 2-halo term, while for the lower redshift (Band 2) the 1-halo term emerges at multipoles larger than about 500.  The cross-correlation signal due to astrophysical sources is dominated by mAGN, SFG and BL Lacs, with the latter being less important for the low-redshift emission of Band 2. This fact arises from an interplay between the redshift dependence and the size of the gamma-ray emission of each individual class of sources, as shown in Fig.~\ref{fig:Cl-behaviours}. For instance, the dominance of SFG and mAGN for the signal coming from $z<0.5$ is traced to the fact that BL Lacs have a suppressed contribution at low redshift.

For the nominal DM case shown here, the DM signal is subdominant by a factor of 10 to 50 for Band 1 and improves to become of a factor of 3 to 5 smaller than the signal from the dominant classes of astrophysical sources for Band 2. As commented above for the behaviour of the window functions, this is due to the fact that the DM unresolved emission is peaked at very low redshift, contrarily to the emission from unresolved astrophysical sources. The comparison between the left and right panels in Figs.~\ref{fig:cross-meerkat}, \ref{fig:cross-ska1}, and \ref{fig:cross-ska2} therefore suggests that higher redshifts (i.e.\ lower detected radio frequencies) are better suited  to pinpoint the contribution from astrophysical sources and then the low redshift investigation can be focussed on the search of a DM signal. Detected frequencies in the range between 950 MHz and the rest-frame 21-cm line frequency are the most promising for a DM search.

Together with the signal predictions, Figs.~\ref{fig:cross-meerkat}, \ref{fig:cross-ska1}, and \ref{fig:cross-ska2} also show the expected uncertainty on the signal. The error estimates are determined according to Eq.~(\ref{eq:deltaClcross}). In the error budget, the cross-correlation signal is always largely subdominant as compared to the two auto-correlation terms. The contribution due to the gamma-ray auto-correlation is largely dominated by the noise term, as can be seen by comparing the noise listed in Table~\ref{tab:noise_gamma} with the gamma-ray auto-correlation signal shown in the right panel of Fig.~\ref{fig:gamma}. The error on the cross-correlation can be thus approximated by:
%, while for the other terms the dominate dependence is:
%
\begin{equation}
(\Delta C_l^{\HI\gamma})^2 \simeq \frac{1}{(2l+1) f_{\rm sky}} \left[ \frac{N^\gamma}{\left(B^\gamma_l\right)^2}  \times \left(C_l^{\HI-\HI} + \frac{N^\HI}{\left(B^\HI_l\right)^2} \right)  \right].
\label{eq:deltaClcross-approx}
\end{equation}

\begin{figure}[t]\centering
\includegraphics[width=0.49\textwidth]{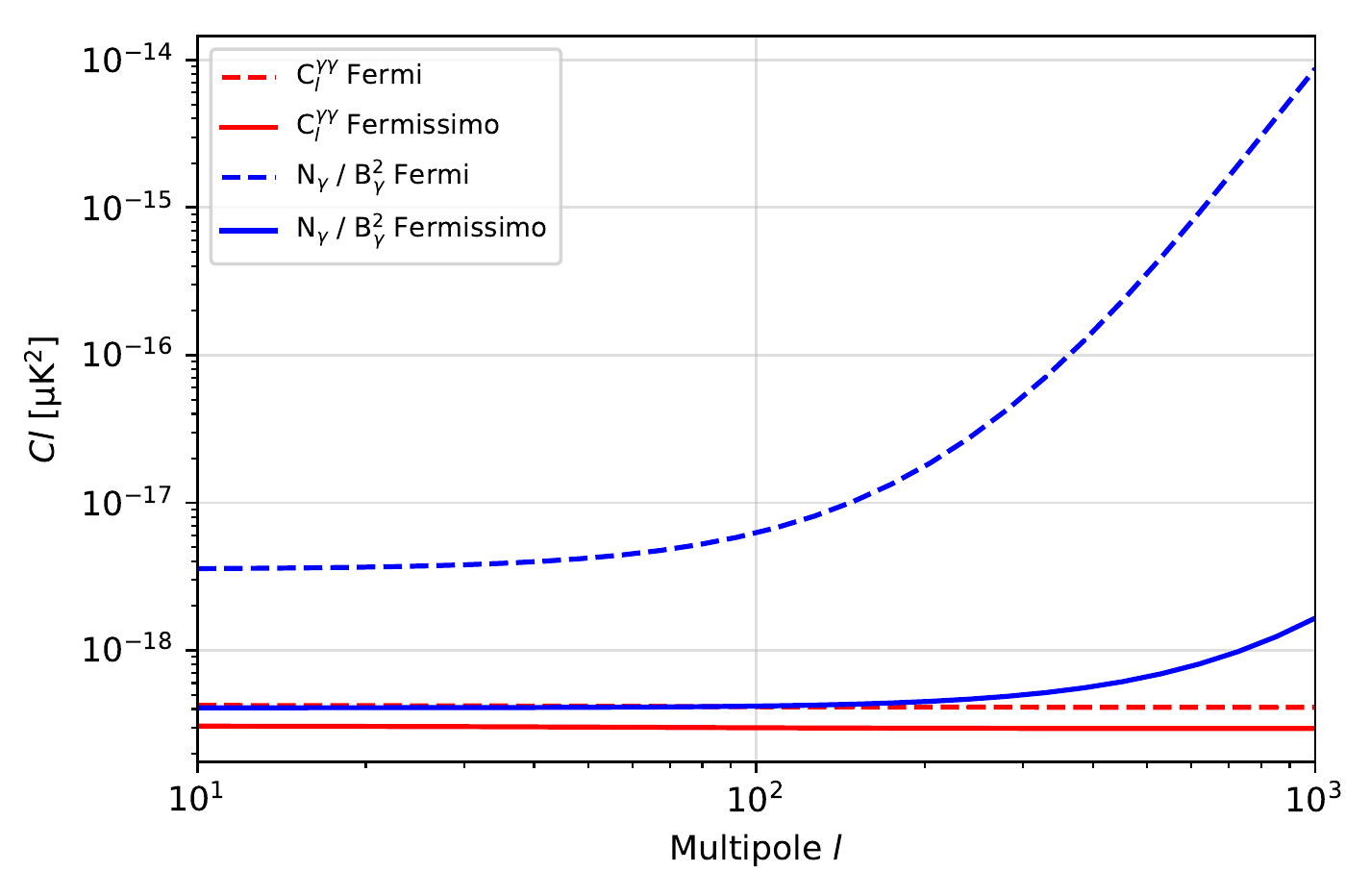}
\includegraphics[width=0.49\textwidth]{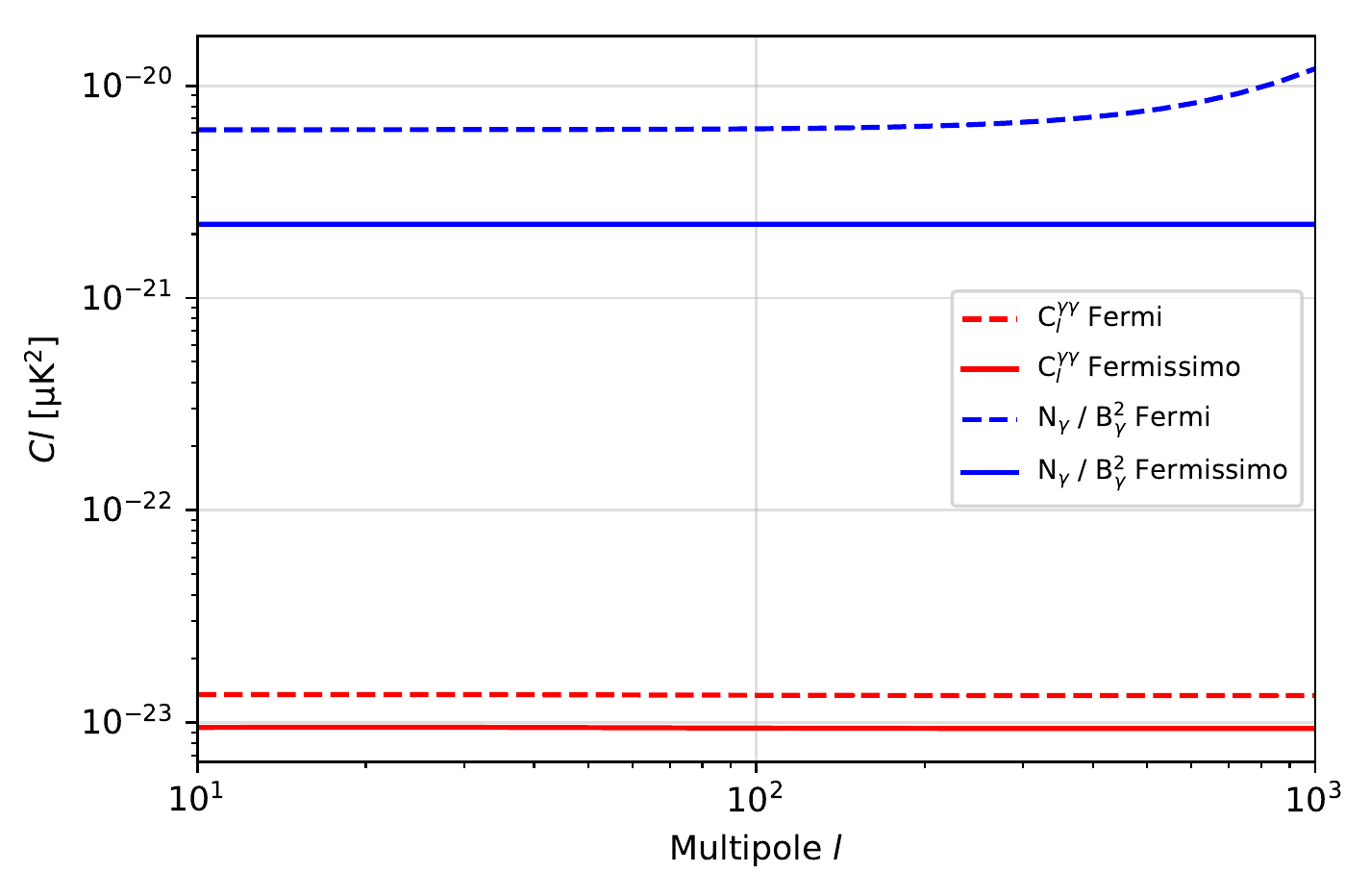}
\caption{Terms contributing to the Gaussian estimate of the variance in the cross-correlation signal and arising from the gamma-ray auto-correlation signals $C_l^{\gamma\gamma}$ and its noise $N^\gamma$. The left panel refers to the energy bin number 2, while the right panel to the energy bin number 10, as reported in Table~\ref{tab:noise_gamma}. The different lines refer to the two contributions for different gamma-ray telescope specifications, as stated in the inset box.} 
\label{fig:error_fermi}
\end{figure}

\begin{figure}[t]\centering
\includegraphics[width=0.49\textwidth]{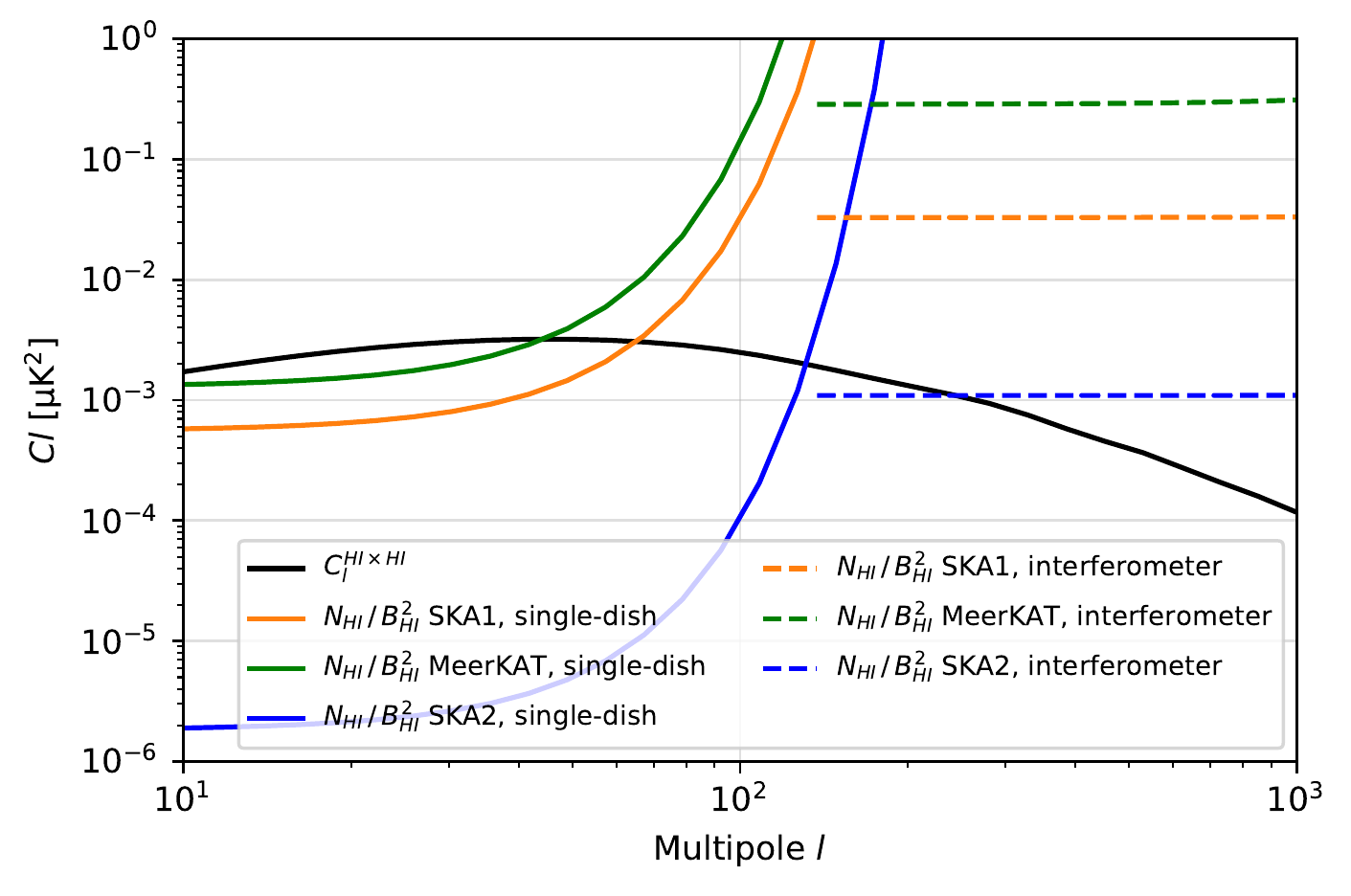}
\includegraphics[width=0.49\textwidth]{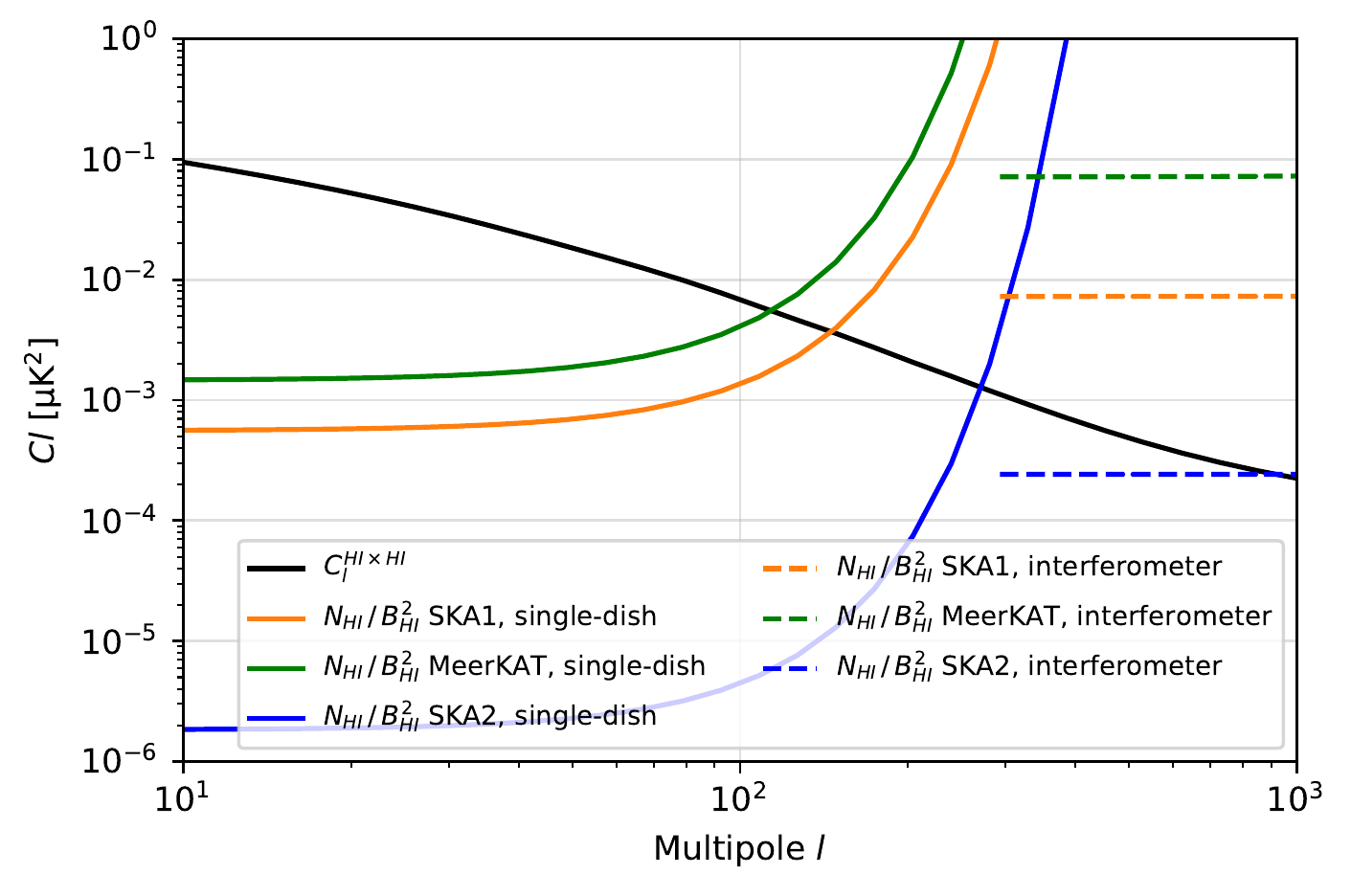}
\caption{Terms contributing to the Gaussian estimate of the variance in the cross-correlation signal and arising from the \hi\, auto-correlation signals $C_l^{\rm HI-HI}$ and its noise $N_{HI}$. The left panel refers to SKA Band 1, while the right panel to SKA Band 2, as reported in Table~\ref{tab:specifiche_IM}. The different lines refer to the two contributions for different radio telescope specifications and configurations, as stated in the inset box.} 
\label{fig:error_ska}
\end{figure}

Fig.~\ref{fig:error_fermi} details the contribution to the variance for a low (bin 1) and high (bin 9) energy bin, as a function of the multipole. The upturn of the noise term for the low energy bin is due to the  detector beam function $B^\gamma_l$, which is suppressed for multipoles larger than 100 at low energies. For larger energies the suppression becomes less relevant for the multipole range adopted here. For the \hi\ intensity mapping,  Fig.~\ref{fig:error_ska} shows that the noise of the single-dish configuration blows up for multipoles of the order of 80 to 100, similar to \fermi\ at low photon energies, with \fermi\ resolution at high energies being better than the one obtained for the radio telescope in single-dish configuration: in this case, the range of multipoles which bring information to the cross-correlation signal is limited by the single-dish resolution. However, in the interferometric configuration, the excellent angular resolution provides a large gain over the single-dish configuration for multipoles larger than $l_{\rm cut}$, i.e.\ 120 for Band 1 and 250 for Band 2. We therefore derive our results for two different configurations for the radio telescopes: single-dish and a combination of single-dish and interferometer, which takes into account the best between the two noise terms. For SKA2, Fig.~\ref{fig:error_ska} also shows that the error budget in Band 2 is (almost) always dominated by the \hi\  auto-correlation term up to large mutipoles, and therefore not limited by the noise term in this case.
From this discussion on the behaviour of the different sources of error, we also determine the maximal value of the multipole over which we focus our analysis: we adopt $l_{\rm max} = 1000$, which is also consistent with the analysis of other types of correlations performed with \fermi\ data (e.g.\, Refs.~\cite{Ammazzalorso:2018evf,Ackermann:2018wlo}).

This discussion on the properties of the different terms entering the error determination helps in understanding the behaviour of the error bars of the cross-correlation signal shown in Figs.~\ref{fig:cross-meerkat}, \ref{fig:cross-ska1} and \ref{fig:cross-ska2},  which refer to the combined dish+interferometer case.  At low multipoles, the error is large due to the low number of modes available in the measurement of the APS, while for high multipoles the error increases due to the size of the \fermi\ and MeerKAT or SKA beams.  However, for all configurations there are windows in multipole where the auto-correlation signal is potentially measurable. The comparison among Figs.~\ref{fig:cross-meerkat}, \ref{fig:cross-ska1} and \ref{fig:cross-ska2} shows the evolution and improvement that can be obtained by progressing from MeerKAT to SKA1 to SKA2.

To determine whether a cross-correlation signal is detectable, we adopt a signal-to-noise ratio (SNR) defined as:
\begin{equation}
{\rm SNR} ^2= \sum_{l,a,r} \left(\frac{C_{l,ar}^{\HI\gamma_\star}}{\Delta C_{l,ar}^{\HI\gamma_\star}}, \right)^2
\end{equation}
where $a$ denotes the energy bin and $r$ the redshift bin. The sum extends on the $N = N_{\rm multipole} \times N_{\rm energy} \times N_{\rm redshift}$ bins. $N_{\rm multipole}  = l_{\rm max} -l_{\rm min}$, for which we adopt $l_{\rm min} = 10$ and $l_{\rm max} = 1000$, as discussed above. $N_{\rm energy} = 12$ (those of Table~\ref{tab:noise_gamma}). $N_{\rm redshift} = 1$ for the analyses in the bands reported in Table~\ref{tab:specifiche_IM} (we have investigated also
a tomographic redshift binning in each band, with $N_{\rm redshift}= 5, 10$, by obtaining only marginal improvements over the results reported here for the single redshift bin). We perform the analysis on the \textit{astrophysical signal $\gamma_\star$ only}, in order to assess the potential of the cross-correlation technique to probe the UGRB independently on any assumption on the presence and size of a DM contribution. Consequences for DM will be discussed right after.

The results are shown in Table~\ref{tab:SNR}, where we see that a hint for the presence of the cross-correlation between the 21-cm brightness temperature and the UGRB is already possible with MeerKAT combined with a statistics of \fermi\ data comparable to the one already available: a SNR of 3.6/3.7 is in fact predicted for both the single-dish and combined (dish+interferometer) configurations, in both redshift bands. With SKA1, a SNR in excess of 5 can be obtained for both configurations in Band2. With SKA2, both redshift bands can allow a clear identification of the signal, with a SNR ranging from 6.7 to 8.2. 

\begin{table}[t]
\centering
\begin{tabular}{llcc}
 \hline
~&~&  Single-dish & Dish+Interferometer \\
\hline
MeerKAT & L-band & 3.6 & 3.6 \\
~& UHF-band & 3.7 & 3.7 \\
\hline
SKA-1 & Band 1 & 4.5 & 4.6 \\
~& Band 2 & 5.7 & 5.7 \\
\hline
SKA-2 & Band 1 & 7.1 & 8.2 \\
~& Band 2 & 6.7 & 7.0 \\
\hline
\hline
\end{tabular}
\caption{Forecast of the signal-to-noise ratio SNR  expected for the cross-correlation between \hi\ intensity mapping and gamma rays emission from astrophysical sources, for different radio telescope configurations combined with \fermi.}
\label{tab:SNR}
\end{table}

Having assessed that a signal is indeed potentially identifiable, we now turn to investigate what kind of bounds on the DM properties this cross-correlation technique can lead to. To determine whether a DM signal can be visible on top of the astrophysical signal, we perform a test on a null hypothesis (presence of the astrophysical signal only) vs. the alternative hypothesis where the astrophysical sources and DM gamma-ray emission are both present. We adopt the statistics:
\begin{equation}
\Delta \chi^2 = \sum_{l,a,r} \left(\frac{C_{l,ar}^{\HI\gamma_{\star +DM}}}{\Delta C_{l,ar}^{\HI\gamma_{\rm \rm \star+DM}}}\right)^2 - \sum_{l,a,r} \left(\frac{C_{l,ar}^{\HI\gamma_\star}}{\Delta C_{l,ar}^{\HI\gamma_\star}}\right)^2,
\end{equation}
where $\gamma_{\rm \star +DM}$ refers to the signal coming from both astrophysical sources and DM, while $\gamma_\star$ is the astrophysical signal only. We perform a raster scan of the DM parameter space over the DM mass: in this case, for each DM mass the free parameter is the annihilation cross-section. In this way, the adopted statistics is distributed as a $\chi^2$ with 1 degree of freedom. We determine the level at which the cross-correlation technique can set a bound on the DM annihilation cross-section at the $2\sigma$ level, i.e.\ we determine the values of $\sigmav$ where $\Delta\chi^2 = 4$.  

\begin{figure}[t]
\centering
\includegraphics[width=0.90\textwidth]{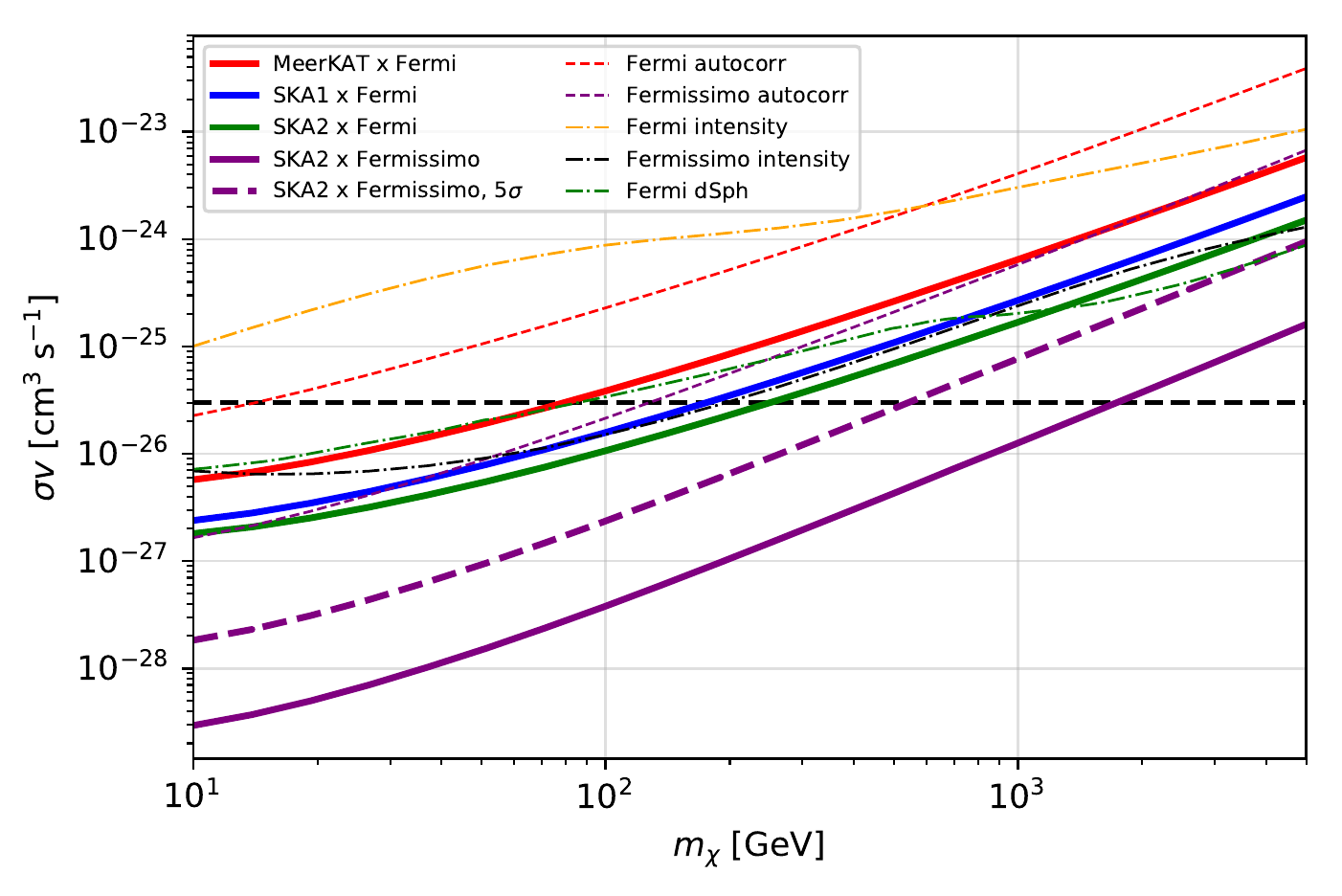}
\caption{Forecast for the bounds on the DM particle properties (mass $m_\chi$ and annihilation rate $\sigmav$) attainable from the study of the cross-correlation between \hi\ intensity mapping and the unresolved component of the gamma-ray background, for different observational set-ups. The annihilation channel is $\bar b b$. Solid lines refer to the 95\% C.L. expected upper bounds on $\sigmav$ for each mass, while the dotted line refers to the region of parameter space where a cross-correlation signal can be detected at the $5\sigma$ C.L. in the configuration SKA2 $\times$ {\it Fermissimo}. The horizontal dashed line outlines the cross-section value required by a massive DM particle to be the dominant DM component. The red  (purple) dashed lines denote the bounds that can be obtained from the gamma-ray autocorrelation angular power spectrum measured by the {\it Fermi}-LAT \cite{Ackermann:2018wlo}  and forecasting the reach of {\it Fermissimo}. The yellow dot-dashed line denotes the bound that can be obtained from the gamma-ray UGRB average flux measured by the {\it Fermi}-LAT \cite{2015IGRB}. The black dot-dashed line is the predicted sensitivity of {\it Fermissimo} derived for the gamma-ray UGRB average flux. The green dot-dashed line shows the current bound on dark matter obtained from dwarf spheroidal galaxies in Ref. \cite{Fermi-LAT:2016uux}.}
\label{fig:bound}
\end{figure}

The results are shown in Fig.~\ref{fig:bound} for the combination of \fermi\ with MeerKAT (first from the top, red curve), \fermi\ with SKA1 (second from the top, red), \fermi\ with SKA2 (third from the top, green). For all cases, we have considered the combined dish+interferometer configuration and they all refer to the lower-redshift band (L-band for MeerKAT and Band 2 for SKA), which is the one more promising for investigating DM since the window function for unresolved gamma rays is prominently peaked at low redshift, while for astrophysical sources it has a peak at intermediate redshift, as shown in Fig.~\ref{fig:bias}. The bounds attainable with Band 2 are a factor of 5 to 10 less constraining, and are therefore not reported here. The plot shows also the bounds that can be obtained from the gamma-ray autocorrelation APS by using {\it Fermi}-LAT data and forecasting the reach of {\it Fermissimo}.  We notice that the cross-correlation method allows to obtain bounds one order of magnitude stronger than those obtainable with the auto-correlation APS when the {\it Fermi}-LAT setup is considered, and potentially a factor of 50 stronger for the {\it Fermissimo} configuration. The same figure reports also the bound that can be obtained from the gamma-ray UGRB average flux measured by the {\it Fermi}-LAT \cite{2015IGRB} (i.e. by using the mean ``isotropic'' unresolved gamma-ray intensity, without resorting to its fluctuations) and the corresponding predicted sensitivity  of {\it Fermissimo}.  The bound from {\it Fermi}-LAT takes into account uncertainties from galactic foreground subtraction, as quoted in  Ref. \cite{2015IGRB}. The prediction for {\it Fermissimo} is done under the same assumptions adopted for the determination of the auto- and cross- correlation sensitivities.
Finally, even though this does not refer to UGRB emission, Fig.~\ref{fig:bound} shows the current bound on dark matter obtained from dwarf spheroidal galaxies obtained by the Fermi-LAT Collaboration \cite{Ackermann:2015zua, Fermi-LAT:2016uux}.

The best obtainable bounds for MeerKAT are deeper in the parameter space as compared to most of the bounds already achieved by cross-correlating gamma rays with galaxies \cite{Ando:2013xwa,Regis2015,Cuoco2015,Shirasaki:2015nqp,Ammazzalorso:2018evf}, clusters of galaxies \cite{Tan:2019gmb} and cosmic shear \cite{Shirasaki2014,Troster:2016sgf,Shirasaki:2016kol,Shirasaki:2018dkz,Ammazzalorso:2019wyr}. SKA1 can improve on the bounds by an additional factor of 4 as compared to MeerKAT, and test a DM particle with thermal annihilation cross-sections for masses up to 130 GeV. SKA2 can further explore the thermal DM particle up to masses of 200 GeV.

For MeerKAT and SKA specifications similar to those reported in Table~\ref{tab:specifiche_IM}, what discussed above almost sets the limit of what can be attainable on the particle DM bounds by performing the cross-correlation between the \hi\ intensity mapping and the unresolved component of the gamma-ray background with \fermi. We have focussed our analysis on 8 years of data taking for the \fermi: it is foreseeable that by the time SKA1 will provide intensity mapping data, \fermi\ will have provided about 50\% more data. This would allow to slightly improve the predicted bounds shown in Fig.~\ref{fig:bound}. However, a  leap in the exploration of the DM parameter space would require a new generation of gamma-ray detectors. In fact, we have shown above that a limiting factor is represented by the gamma-ray angular resolution, which boosts the gamma rays noise term in Eq.~(\ref{eq:deltaClcross-approx}) for multipoles above a few hundreds. This occurs also for MeerKAT and SKA1 in single-dish configuration, with a mild improvement for high multipoles by the addition of interferometric data. However, this situation can be largely overcome by SKA2, for which the noise term becomes subdominant as compared to the cosmic variance one, especially for low redshift. In order to investigate the potentiality of the cross-correlation signal in investigating particle DM, we examine the capabilities of a future gamma-ray detector set-up endowed with the following characteristics. First, we assume the  exposure of the detector to be larger by a factor of 2 as compared with the current \fermi\ specification adopted here: this implies that the limiting sensitivity to unresolved sources $L_{\rm sens}$ is scaled down by the square-root of the increase in the exposure, i.e.\ by a factor 1.4, and the signal coming from unresolved astrophysical sources gets slightly diminished (while the DM signal remains unchanged). Second, we assume that the detector point-spread-function can be improved, and we adopt the same behaviour of the beam function expressed in Eqs.~(\ref{eq:beam1}) and (\ref{eq:beam2}) but with a better angular resolution \cite{Topchiev:2017xfp}, viz.\ :
\begin{equation}
\sigma_0(E) = \alpha_\sigma \times \sigma_0(E_{\rm ref})  \times (E/E_{\rm ref})^{-0.95} + 0.001,
\end{equation}
for which we assume for definiteness $\alpha_\sigma = 0.2$. Finally, thanks to the better angular resolution, which means smaller mask, we adopt a larger sky-fraction coverage of $f_{\rm sky} = 0.8$, which allows to slightly reduce the impact of noise. For definiteness, we work with the same energy bins of Table~\ref{tab:noise_gamma}: this allows us to directly rescale the noise estimate in terms of the adopted changes in exposure and $f_{\rm sky}$, since the noise can be determined as \cite{Ackermann:2012uf}: $N^\gamma = N_{\rm photons}/(A^2 \, \Omega_{\rm obs})$ where $A$ is the exposure, $N_{\rm photons}$ is the number of photons ($N_{\rm photons}=I\,\delta E\,A\,\Omega_{\rm obs}$) in the observed solid angle $\Omega_{\rm obs} = 4\pi f_{\rm sky}$. Thus $N^\gamma$ scales as $\sim 1/A$. We call this set-up \textit{Fermissimo} (similarly to what was done in Ref.~\cite{Camera:2014rja} when forecasting the cross-correlation signal with cosmic shear) and we perform the analysis for its combination with SKA2.

\begin{figure}[t]
\centering
\includegraphics[width=0.49\textwidth]{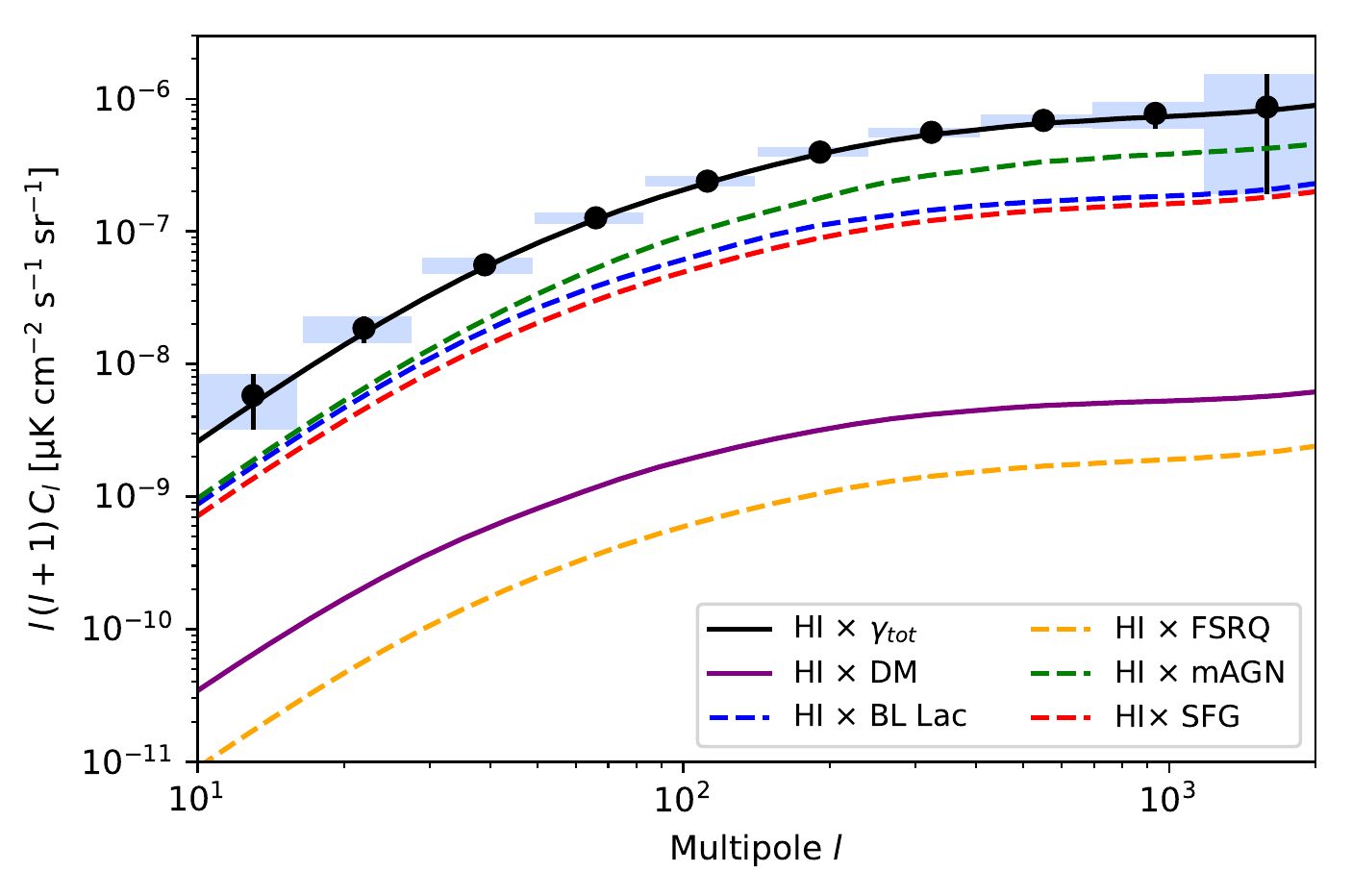}
\includegraphics[width=0.49\textwidth]{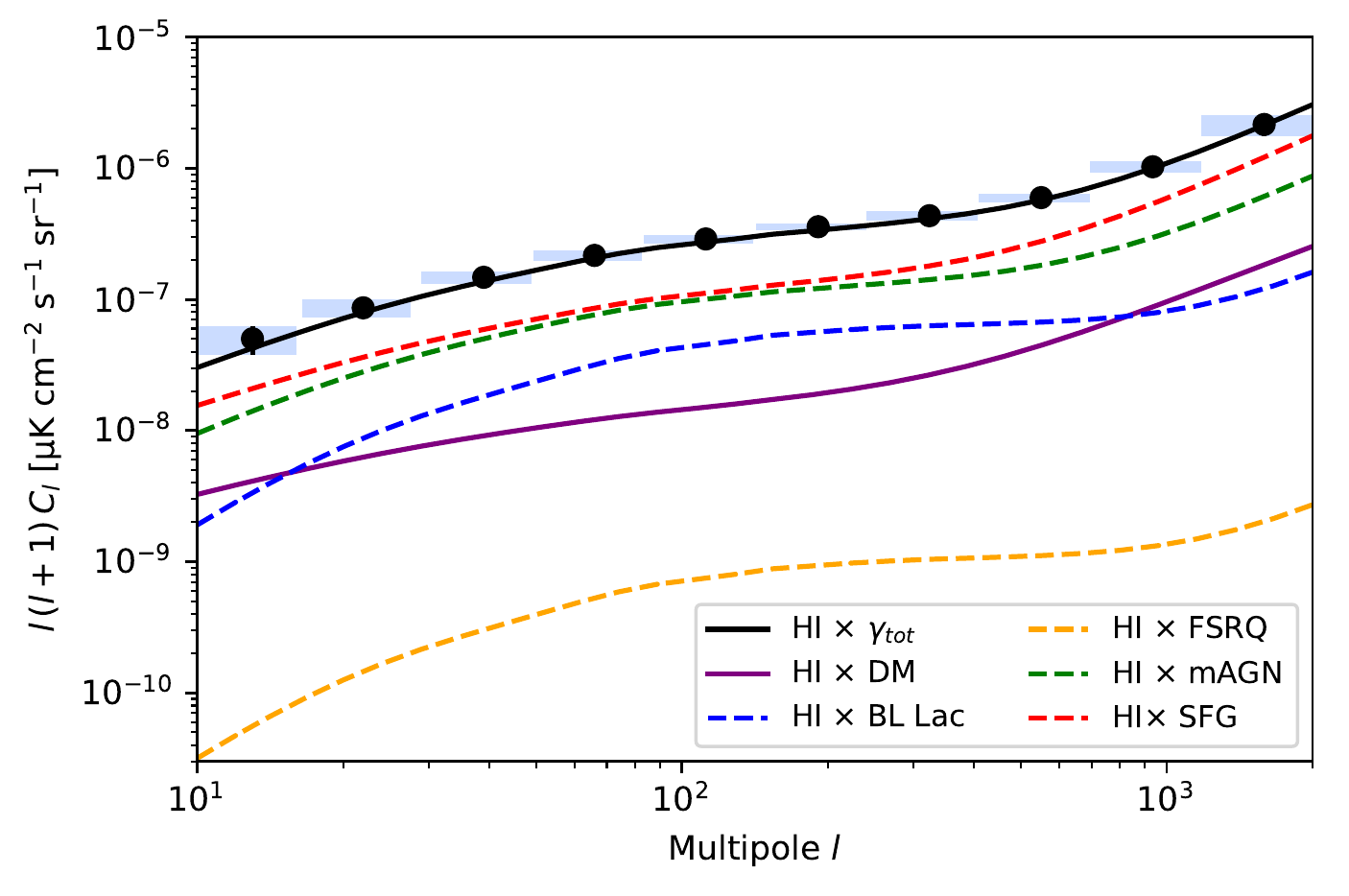}
\caption{The same as in Fig.~\ref{fig:cross-meerkat} (except that the multipole scale is here extended up to $l=2000$), for SKA2 $\times$ {\it Fermissimo}. The left panel refers to the higher redshift Band 1, the right panel to the lower redshift Band 2.}
\label{fig:cross-ska2fermissimo}
\end{figure}

The impact of the improved gamma-ray angular resolution is clearly visible in Fig.~\ref{fig:cross-ska2fermissimo}, as compared to Fig.~\ref{fig:cross-ska2}, and reflects into a significant reduction of the error bars for multipoles beyond 300 and moreover allows to extend the analysis to larger multipoles (for which we set $l_{\rm max} = 2000$) . The ensuing implications on the DM investigation are shown in the lowest (purple) curve in Fig.~\ref{fig:bound}. The 95\% C.L.  bound is shifted down by a factor between 10 and 60, as compared to the bound arising from SKA2 in combination with {\it Fermi}. The whole mass range of a thermal DM particle up to the TeV scale can be tested. On the same plot we also show the $5\sigma$ detection reach (dotted line), which allows detection of a particle DM with thermal cross-section up to masses of 400 GeV.

\section{Conclusions}
\label{sec:conclusions}
In this paper we have explored the idea to use the 21-cm line of neutral hydrogen as a gravitational tracer of the matter distribution in the Universe to investigate the nature of the UGRB, through the adoption of the cross-correlation technique. Since cosmological gamma-ray emission dominantly occurs in the same cosmic structures hosting neutral hydrogen, a positive correlation is in fact expected.

We have quantified the size of this effect by investigating the small fluctuations due to the inhomogeneous distribution of matter in the late Universe. The large-scale structure distribution of matter in the Universe, from one side induces fluctuations in the 21-cm brightness temperature emission of neutral hydrogen, on the other side produces fluctuations in the unresolved component of the gamma-ray background: these fluctuations are due to either astrophysical sources hosted by those cosmic structures, or to DM in the form of particles which annihilate and produce gamma rays through their annihilation products. 

We have studied the angular power spectrum of cross-correlation between these two kinds of fluctuations and found that data from future campaigns of neutral hydrogen intensity mapping measurements combined with the current sensitivity of the \fermi\ gamma-ray telescope, have the capability to detect the cross-correlation signal. We obtained that the combination of MeerKAT with the current \fermi\ statistics can provide a first hint of the cross-correlation signal due to astrophysical sources, with a signal-to-noise ratio of 3.7. We have then performed forecasts for SKA Phase 1 and Phase 2, again combined with the current sensitivity of \fermi: in these cases, the signal-to-noise ratio is predicted to increase to 5.7 and 8.2, respectively for SKA1 and SKA2.

Having found potential detectability of the signal originated by astrophysical sources, we have investigated the capabilities of this technique to probe particle DM signatures. We predict that the attainable bounds on DM properties are quite competitive with those obtained from other techniques able to explore the unresolved side of the gamma-ray background, like the cross-correlation of gamma radiation with galaxy \cite{Ando:2013xwa,Regis2015,Cuoco2015,Shirasaki:2015nqp,Ammazzalorso:2018evf},
 and galaxy cluster-catalogues \cite{Tan:2019gmb}, CMB \cite{Feng2016} or the cosmic shear \cite{Shirasaki2014,Troster:2016sgf,Shirasaki:2016kol,Shirasaki:2018dkz,Ammazzalorso:2019wyr}. The enhanced capabilities of SKA Phase 2, combined with a future generation gamma-ray telescope with improved specifications will instead allow to investigate the whole mass window for a thermal WIMP up to the TeV scale, with a $5\sigma$ detection possible for  DM masses up to 400 GeV. In order to obtain this enhanced sensitivity, the main requirement for a future gamma-ray telescope is an improved angular resolution, which would allow to better exploit the excellent angular resolution of the interferometric configuration of SKA2 in the determination of the cross-correlation signal. On the other hand, an exposure similar or slightly larger than the one currently attained by \fermi\ would be adequate.

\acknowledgments
This work  is supported by:  `Departments of Excellence 2018-2022' grant awarded by the Italian Ministry of Education, University and Research (\textsc{miur}) L.\ 232/2016; Research grant `The Anisotropic Dark
Universe' No.\ CSTO161409, funded by Compagnia di Sanpaolo and University of Turin; Research grant TAsP (Theoretical Astroparticle Physics) funded by \textsc{infn}; Research grant `The Dark Universe: A Synergic Multimessenger Approach' No.\ 2017X7X85K funded by \textsc{miur}; 
Research grant `From Darklight to Dark Matter: understanding the galaxy/ matter  connection to measure the Universe' No.\ 20179P3PKJ funded by \textsc{miur}; Research grant `Deciphering the high-energy sky via cross correlation' funded by the agreement ASI-INAF n.2017-14-H.0. 
SC is supported by \textsc{miur} through the Rita Levi Montalcini project `\textsc{prometheus} -- Probing and Relating Observables with Multi-wavelength Experiments To Help Enlightening the Universe's Structure.


\end{document}